\documentclass[a4paper,11pt]{article}
\pdfoutput=1

\usepackage{jcappub}

\usepackage[T1]{fontenc}
\usepackage[utf8]{inputenc} 
\usepackage[english]{babel}
\usepackage{float}
\usepackage{subfigure}

\newcommand{\be}{\begin{equation}}
\newcommand{\ee}{\end{equation}}
\newcommand{\bea}{\begin{eqnarray}\displaystyle}
\newcommand{\eea}{\end{eqnarray}}

\newcommand{\Nside}{N_\text{side}}
\newcommand{\lmax}{{\ell_\text{max}}}

\newcommand{\LCDM}{$\Lambda$CDM}
\renewcommand{\deg}{^\circ}
\newcommand{\plotW}{0.48\textwidth}

\DeclareMathOperator{\E}{E}
\DeclareMathOperator{\Var}{Var}

\title{Skewness and Kurtosis as Indicators of Non-Gaussianity in Galactic Foreground Maps}

\author[a,b]{Assaf Ben-David}
\author[b]{, Sebastian von Hausegger}
\author[a]{and {Andrew D. Jackson}}

\affiliation[a]{Niels Bohr International Academy, The Niels Bohr Institute, University of Copenhagen, Blegdamsvej~17, DK-2100 Copenhagen~\O, Denmark}
\affiliation[b]{Discovery Center, The Niels Bohr Institute, University of Copenhagen, Blegdamsvej~17, DK-2100 Copenhagen~\O, Denmark}

\emailAdd{bendavid@nbi.dk}
\emailAdd{s.vonhausegger@nbi.dk}

\abstract{Observational cosmology is entering an era in which high precision will be required in both measurement and data anal\-y\-sis. Accuracy, however, can only be achieved with a thorough understanding of potential sources of contamination from foreground effects.  Our primary focus will be on non-Gaussian effects in foregrounds.  This issue will be crucial for coming experiments to determine $B$-mode polarization.  We propose a novel method for investigating a data set in terms of skewness and kurtosis in locally defined regions that collectively cover the entire sky.  The method is demonstrated on two sky maps: (i) the SMICA map of Cosmic Microwave Background fluctuations provided by the Planck Collaboration and (ii) a version of the Haslam map at 408~MHz that describes synchrotron radiation. We find that skewness and kurtosis can be evaluated in combination to reveal local physical information.  In the present case, we demonstrate that the statistical properties of both maps in small local regions are predominantly Gaussian.  This result was expected for the SMICA map.  It is surprising that it also applies for the Haslam map given its evident large scale non-Gaussianity.  The approach described here has a generality and flexibility that should make it useful in a variety of astrophysical and cosmological contexts.}

\begin{document}
\maketitle
\flushbottom

\section{Introduction} 
\label{sec:introduction}

The release of the Planck Cosmic Microwave Background (CMB) temperature fluctuations data~\cite{Adam:2015rua} represents a landmark 
in observational cosmology.  The standard cosmological model (\LCDM) of an isotropic and homogeneous Universe, emerging after an inflationary era and today containing mainly dark energy and cold dark matter, has proven to be highly successful in describing the CMB data.  Its six parameters have been remarkably well determined and constrained~\cite{Ade:2015xua}.   Although there may be more that can be learned from CMB temperature data, the community is now turning its attention to measurements of CMB polarization signals with the specific aim of searching for signs of primordial $B$-mode polarization. This task presents some major challenges. This signal, if it exists, will be extremely faint relative to foreground effects and instrumental noise.  While our understanding of the properties of various Galactic foregrounds contaminating the CMB signals has been sufficient for the temperature anal\-y\-ses, a much better level of measurement, understanding and modelling will be required in order to clean the CMB polarization sky maps to the level necessary for primordial $B$-mode detection.  It is therefore essential to study the characteristics of foreground sky maps in detail, including their statistical properties.  Synchrotron radiation is one of the most important of these foregrounds, and its temperature power spectrum has been the subject of several studies, e.g.~\cite{Mertsch:2013pua,Regis:2011ji}.  Issues 
related to the statistical properties of this foreground remain to be investigated. (Regarding $B$-mode maps, see~\cite{PhysRevLett.113.191303})\\

In this work we present a general approach for studying sky maps by investigating their local one-point distribution in small patches. This approach has the advantage of being blind to the global structure of the map.  An obvious feature of foreground maps is that they reflect structures in our Galaxy.  It is unreasonable to expect that its myriad distinct, specific and complex features can be regarded as a whole suitable for a global anal\-y\-sis with statistical tools.  On the other hand, simple statistical measures can provide useful tools for an\-a\-lyz\-ing sufficiently small local patches of these maps.\footnote{These patches, which will be defined more precisely below, are locally defined non-overlapping regions that collectively cover the entire sky.}  This is simply an example of the fact that the properties of data sets are often scale-dependent. By selecting an appropriate patch size, it is possible to focus on a specific scale of interest and thus to add flexibility to the anal\-y\-sis.  Working with patches also enables us to identify regions of the sky with anomalous properties such as non-Gaussian regions of the CMB map and Gaussian regions of foreground maps.  In addition, the characterization of local foreground properties can be an important aid in the design of experiments that wish to focus on small regions of the sky with favorable local properties.  Given the large scale structures expected in foreground maps, it is natural to consider the  dimensionless moments (i.e., moments that are independent of the local mean and variance) one patch at a time.  Thus, in this work we investigate the use of the skewness and kurtosis as statistical probes. Certainly, many other statistical tests can and should be considered in the study of foreground sky maps, and our focus on the skewness and kurtosis will serve as a simple example of one such test. We hope that a broader variety of statistical tests employing, e.g., Minkowski functionals~\cite{Ade:2015hxq} and the Kullback--Leibler divergence~\cite{Ben-David:2015sia}, will improve our understanding of the various Galactic foregrounds and that they will help pave the way to high-precision anal\-y\-ses of CMB polarization patterns.\\

We will demonstrate our method of anal\-y\-sis using two sky maps with quite different properties.  Since our primary focus is the investigation of foreground maps, we will consider the 408~MHz radio map of Haslam et al.~\cite{Haslam:1982zz} in a recent reprocessed version~\cite{Remazeilles:2014mba}.  This sky map, dominated by Galactic synchrotron emission, is used to model the synchrotron foreground contamination of the CMB signal at higher frequencies~\cite{Adam:2015wua}.  It is considered to be non-Gaussian in nature and serves as a typical example of a foreground component relevant for CMB anal\-y\-sis. While its power on various scales has been studied and is understood (see, e.g.~\cite{Mertsch:2013pua}), our anal\-y\-sis of the one-point distribution and its Gaussianity can be performed `blindly', without invoking knowledge of the physics behind Galactic synchrotron emission. Before considering the Haslam map, however, we will first perform a similar investigation of the Planck 2015 SMICA sky map of CMB temperature fluctuations~\cite{Adam:2015tpy}. This map and its predecessors have been studied in considerable detail in recent years, and its properties are well-understood.  It is known in particular that it can be regarded as consistent with a realization of a statistically homogeneous and isotropic random Gaussian process.  Since the main motivation for our work is the study of sky maps, which naturally have the topology of a sphere, our examples and discussions involving spatial correlations etc.\ will employ the language and notations relevant for a sphere. This is done without loss of generality or relevance to other types of data sets.\\  

The structure of the paper is as follows.  In Section~\ref{sec:skewness_and_kurtosis} we discuss general properties of the skewness and kurtosis of a data set consisting of uncorrelated draws and the complications that arise when the draws are correlated. We also describe our method of analysis. In Section~\ref{sec:analysis_of_sky_maps} we apply our method to the Planck 2015 SMICA map (expected to be Gaussian) and the Haslam map (expected to be far from Gaussian).  Sample regions of the Haslam map will be shown to illustrate some of the features that our method can identify.  These results will be discussed and conclusions drawn in Section~\ref{sec:discussion}.


\newpage

\section{Skewness, kurtosis and non-Gaussianity} 
\label{sec:skewness_and_kurtosis}

\subsection{Definitions and General Properties} 
\label{sub:definitions_and_general_properties}

Consider a data set consisting of $n$ random draws of the variable $x$ on a given distribution.  
The moments of this data set relative to the mean, $\overline{x}$, are given simply as
\be\label{eq:moment_def}
m_k = \frac{1}{n} \sum_i (x_i - \overline{x})^k, 
\ee
where $m_2$ is the usual variance.  For present purposes we will be concerned with the moments 
in various patches of the sky.  Thus, the sum in Eq.~\eqref{eq:moment_def} should extend over all 
pixels in the patch under consideration, and the mean $\overline{x}$ is similarly that of the patch.
For $k > 2$, it is useful to adopt $m_2^{1/2}$ as a unit of 
length in order to obtain moments, $\mu_k = m_k/m_2^{k/2}$, that depend on the shape of the 
underlying distribution but are independent of its scale.  If the distribution in question is Gaussian, 
$\mu_k = 0$ for odd $k$ and $\mu_k = (k-1)!!$ for even $k$ in the limit $n \to \infty$.  For 
studying the non-Gaussianity of a general distribution, it is convenient to measure these moments 
relative to their Gaussian values.  Thus, the skewness and excess kurtosis are defined respectively as 
\be
\gamma_1 = \mu_3 \ \ \text{and}\ \ \gamma_2 = \mu_4 - 3.
\ee
The skewness and excess kurtosis of the elements of our data set have distributions whose 
low moments are known for the special case of Gaussian distributions. Specifically, the mean 
and variance values of $\gamma_1$ are
\bea\label{eq:gamma1}
\E[\gamma_1] &=& 0 \nonumber \\
\Var[\gamma_1] &=& \frac{6(n-2)}{(n+1)(n+3)},
\eea
while for $\gamma_2$ we have
\bea\label{eq:gamma2}
\E[\gamma_2] &=& -\frac{6}{n+1} \nonumber \\
\Var[\gamma_2] &=& \frac{24n(n-2)(n-3)}{(n+1)^2(n+3)(n+5)}.
\eea
Further, each of these distributions is known to be normal in the limit of large $n$.  

Comparison of the distribution of skewness {\em or\/} excess kurtosis resulting from random draws 
on an unknown distribution with such known results can provide a test of non-Gaussianity.  We will 
argue here that it is of potentially greater value to consider the distribution of skewness {\em and\/} 
kurtosis.  Before addressing this point, however, it is important to note that the values of $\gamma_1$ 
and $\gamma_2$ resulting from $n$ random draws on any given distribution are not independent.  
To see this, we consider a single random draw of any size, $n$, on an arbitrary distribution.  Without loss 
of generality, we shift and rescale these numbers to ensure mean $0$ and variance $1$.   Calculate the average 
value of the non-negative function $(x-a)^2 (x-b)^2$ in terms of the moments $m_3$ and $m_4$ for this draw, 
and minimize the result with respect to the real parameters $a$ and $b$.  The result of these operations is the 
Pearson inequality~\cite{Pearson429},
\be \label{eq:pearson}
\gamma_2 \ge \gamma_1^2 - 2.
\ee
Note that the equality is satisfied if and only if the distribution permits only the 
values $x=a$ and $x=b$.  Thus, although stronger inequalities can be constructed 
for special classes of distributions (e.g., unimodal or symmetric unimodal distributions~\cite{Rohatgi1989297,Klaassen2000131}), 
there is no stronger general inequality involving skewness and kurtosis.

As stated above, the distributions of skewness and kurtosis are normal in the large-$n$ limit for the 
special case of a Gaussian.  Noting that their associated variances differ by a factor of $4$ in this 
limit, Jarque and Bera~\cite{Jarque1980255} were led to study the properties of the combination $(\gamma_1^2 + 
\gamma_2^2 /4)$ and to show that its asymptotic behaviour in the large-$n$ limit is precisely 
that of a $\chi^2$-distribution with two degrees of freedom.  In principle, comparison with this 
familiar distribution offers a more accurate measure of Gaussianity than either the skewness or 
kurtosis alone.  In practice, convergence to the asymptotic limit is slow.  This has prompted 
the introduction of ad hoc transformations\cite{DAgostino1970} to render the skewness and kurtosis 
``more Gaussian'' with the aim of improving the accuracy of this combined measure.  The fact that 
these transformations treat skewness and kurtosis as independent means that they cannot 
respect the Pearson inequality and serves as a reminder that their utility is greatest in the region 
of maximum probability ($\gamma_1 = 0$ and $\gamma_2 \to 0$).  

With the aid of Eqs.~\eqref{eq:gamma1} and \eqref{eq:gamma2}, the skewness and kurtosis can be useful 
in deciding whether a given data set is the result of uncorrelated random draws on a Gaussian or a non-Gaussian 
distribution.  They are not a priori suitable for treating systems such as the SMICA map of CMB temperature 
fluctuations where the expectation of uncorrelated random draws in harmonic space necessarily implies the existence 
of correlations in pixel space.  We will address this problem in the next section.


\subsection{Correlated Data} 
\label{sub:correlated_data}

When attempting to analyze the distribution of pixel values of sky maps in small patches, we must deal with the complication that these pixels are expected to be correlated. Maps depicting mainly Galactic foreground emissions are correlated on scales corresponding to the sizes of structures in the Galaxy, which can vary greatly when projected on the sphere.  According to the standard cosmological model, however, maps of CMB temperature fluctuations, such as the Planck SMICA map, are expected to be correlated in pixel space.  Although the harmonic coefficients can be assumed to have been drawn uncorrelated in harmonic space, the transformation to the pixel domain then ensures that the pixels are correlated. Assuming a homogeneous and isotropic distribution based on the angular power spectrum, $C_\ell$, the angular correlation function depends only on the separation angle, $\theta$, between pixels and is given by
\be
C(\theta) = \sum_\ell \frac{2\ell+1}{4\pi} C_\ell P_\ell(\cos\theta),
\ee
where $P_\ell$ are the Legendre polynomials. We can then define the expected angular scale of the correlation, $\theta_\text{c}$, as

\be \label{eq:corr_angle}
\theta_\text{c}^2 = \frac{C(0)}{|C''(0)|}
= \frac{2\sum_\ell (2\ell+1) C_\ell} {\sum_\ell \ell (\ell+1) (2\ell+1) C_\ell}.
\ee
For example, in this work we will consider the SMICA map smoothed with a Gaussian kernel of $20'$ FWHM, which allows us to use only scales up to $\lmax=1000$. If we choose $C_\ell = g_\ell^2 C_\ell^{\Lambda\text{CDM}}$, where $C_\ell^{\Lambda\text{CDM}}$ is the Planck best-fit \LCDM\ power spectrum~\cite{Aghanim:2015xee} and $g_\ell$ is the smoothing kernel, Eq.~\eqref{eq:corr_angle} results in a correlation angle $\theta_\text{c} = 29'$. When considering data in a patch of angular size $\theta_\text{p}$, we should expect that correlations will have a non-negligible effect on the joint distribution of the skewness and excess kurtosis of the data unless $\theta_\text{p} / \theta_\text{c} \gg 1$.  

\begin{figure*}
	\centering
	\subfigure[]{
		\includegraphics[width=0.47\textwidth]{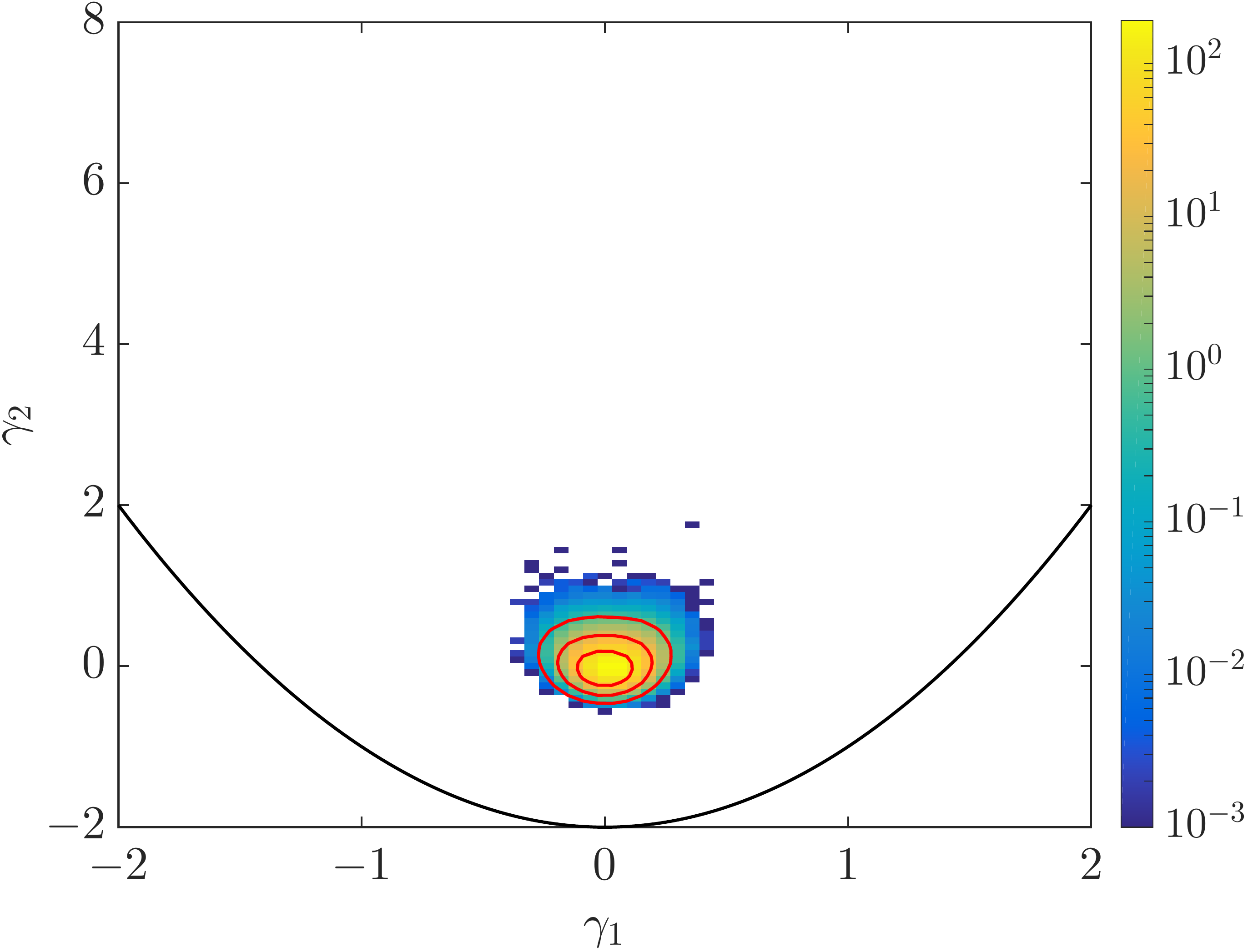}
		\label{subfig:UncorrelatedEnsemble}
	}
	\subfigure[]{
		\includegraphics[width=0.47\textwidth]{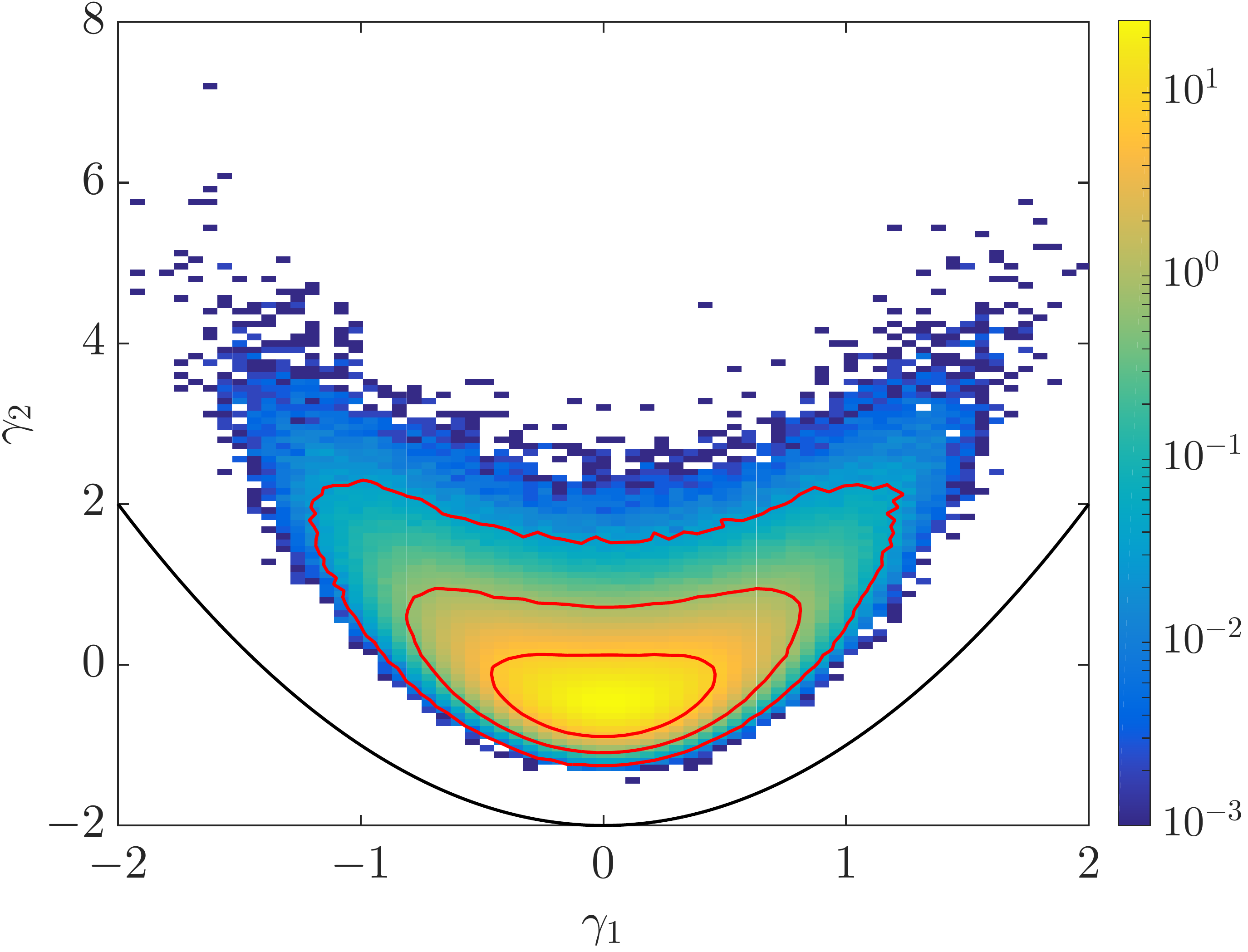}
		\label{subfig:mean_histogram_smica}
	}
	\caption{Mean histogram of skewness and excess kurtosis, calculated on \subref{subfig:UncorrelatedEnsemble}~uncorrelated pixels and \subref{subfig:mean_histogram_smica}~correlated pixels with $\theta_\text{p}/\theta_\text{c} = 7.5$. The colors represent (mean) counts, on a logarithmic scale. Also plotted are the constraint parabola~(\emph{black}) imposed by Pearson's inequality~\eqref{eq:pearson} and the 1--3$\sigma$ contour lines~(\emph{red}). \label{fig:CompareCorrelatedToUncorrelated}}
\end{figure*}

The effect of correlations is readily demonstrated. We use the hierarchical nature of \textsc{HEALPix}\footnote{\url{http://healpix.sourceforge.net}}~\cite{Gorski:2004by} to partition a sky map of resolution $\Nside = 512$ into non-overlapping patches by assigning each pixel to a patch on a grid of $\Nside=16$. This results in $N_\text{p} = 3072$ patches with an angular size of $\theta_\text{p} = 3.7\deg$, each containing 1024 pixels. We calculate the skewness and excess kurtosis of the pixels in each patch and bin them into a bivariate histogram. In order to smooth the statistical fluctuations of the histograms, we draw 1000 realizations of sky maps and calculate the mean of all histograms. We consider two such ensembles.  The first is drawn in pixel space with pixels drawn uncorrelated from a normal distribution.  The second ensemble is drawn in harmonic space, and we use the smoothed angular power spectrum $C_\ell$ mentioned above to draw the uncorrelated harmonic coefficients from a Gaussian distribution.  This results in 
correlated pixels with $\theta_\text{p}/\theta_\text{c} = 7.5$. The two mean histograms are shown in Fig.~\ref{fig:CompareCorrelatedToUncorrelated}.
The difference between the two is striking. Pixel correlations cause the distributions of both $\gamma_1$ and $\gamma_2$ to be much wider, and their interdependence is made more apparent by the fact that the contour lines of the histogram are no longer elliptic.

We therefore require an alternative strategy for analyzing the joint distribution of $\gamma_1$ and $\gamma_2$ on the patches of a given sky map that will allow for  correlations of the pixels.  To this end, we have chosen a straightforward generalization of the strategy --- common in one dimension --- of comparing to an ensemble. Given an ensemble of realizations, we calculate the mean bivariate histogram, as those shown in Fig.~\ref{fig:CompareCorrelatedToUncorrelated}, and use it as a proxy for the two dimensional probability distribution of the moments. This histogram, $h_{ij} = h(\gamma_{1,i}, \gamma_{2,j})$, where $\gamma_{1,i}$ and $\gamma_{2,j}$ are the bins of the moments, can now be used to assign a $p$-value to each point, $(\gamma_1, \gamma_2)$, in the skewness--kurtosis plane in the following way. First, we calculate the value $h(\gamma_1, \gamma_2)$ by interpolation of the discrete histogram $h_{ij}$. We then find the bins comprising the appropriate `tail' of the distribution, i.e.\ the set of all bins with smaller counts, $T = \{(i, j) \mid h_{ij} \le h(\gamma_1, \gamma_2)\}$. Finally, the probability to exceed the given value is calculated as $p = N_\text{p}^{-1} \sum_{(i, j) \in T} h_{ij}$, where the number of patches, $N_\text{p}$, is obviously also the sum of $h_{ij}$ over all bins. This procedure allows us to draw in Fig.~\ref{fig:CompareCorrelatedToUncorrelated} contours representing the 1--3$\sigma$ bounds of the distribution, where we use the standard definition for the normal distribution to convert the language of $p$-values to that of $\sigma$-values, i.e.\ $1\sigma$ corresponds to $p=31.7\%$, etc.

With a $p$-value assigned to each patch, it is possible to identify anomalous regions and to inspect them more closely. However, these $p$-values still depend on the nature of the ensemble of realizations chosen as a basis of comparison. We will utilize this procedure to analyze the SMICA and Haslam maps in the following section.   We note that the assumption of statistical homogeneity and isotropy cannot be justified for foreground maps such as the Haslam map.  While there is thus no guarantee that the strategy outlined here is quantitatively reliable when applied to foreground maps, we believe that it can provide useful qualitative guidance.



\section{Analysis of Sky Maps} 
\label{sec:analysis_of_sky_maps}

\subsection{The SMICA Map} 
\label{sub:the_smica_map}

\begin{figure}
	\centering
	\includegraphics[width=\plotW]{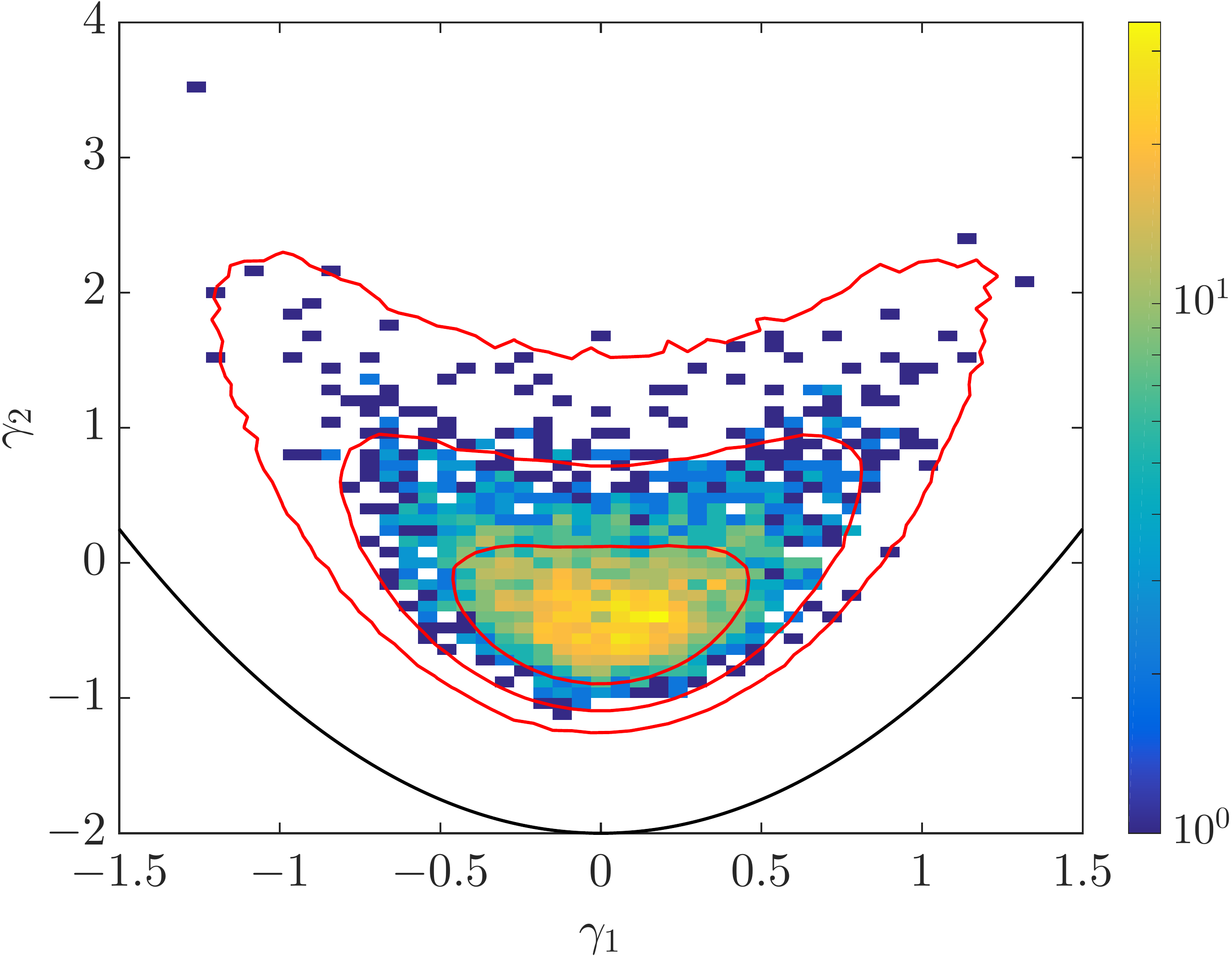}
	\caption{The skewness--kurtosis histogram for the SMICA map. Also plotted are the constraint parabola and the same contour lines as in Fig.~\ref{subfig:mean_histogram_smica}. \label{fig:histogram_smica}}
\end{figure}

We start by analyzing the Planck 2015 SMICA map of CMB temperature fluctuations~\cite{Adam:2015tpy}. 
We degrade the map from its native \textsc{HEALPix} resolution of $\Nside=2048$ to a resolution of $\Nside=512$ after first smoothing it with a Gaussian kernel of $20'$ FWHM\@. Using an $\Nside=16$ grid for the patches as described above, we obtain the skewness--kurtosis of the SMICA patches shown in Fig.~\ref{fig:histogram_smica}.
We note that the choice of smaller patches, e.g.\ by using a grid of $\Nside=32$, would result in larger statistical fluctuations of the moments since each patch would contain fewer pixels.  In addition, the smaller patch would be closer to the correlation angle of the map. While smaller patches could still be used for SMICA, they would not be appropriate for the Haslam map to be analyzed below.

In order to assess the expected distribution of the moments for the SMICA map, we draw a Gaussian ensemble for purposes of comparison. While the SMICA map also contains other effects (such as instrumental noise and foreground residuals) in addition to the dominant Gaussian signal, it was recently shown~\cite{Ben-David:2015sia} that the statistical properties of a Gaussian ensemble are close to those of the Planck full focal plane (FFP) simulations, which include some of these additional effects on sufficiently large scales. We therefore use a Gaussian ensemble here and will show that the statistical properties of the skewness and excess kurtosis of the SMICA map are consistent with it.  Since we wish to ensure that the Gaussian ensemble has the same correlation structure in pixel space as the SMICA map, we use the ensemble of 1000 realizations based on the best-fit \LCDM\ angular power spectrum described in Section~\ref{sub:correlated_data}. The mean histogram for this ensemble was shown in Fig.~\ref{subfig:mean_histogram_smica}, and the 1--3$\sigma$ contour lines calculated based on it are superimposed on the SMICA histogram in Fig.~\ref{fig:histogram_smica}.

\begin{figure*}
	\centering
	\subfigure[]{
		\includegraphics[width=0.47\textwidth]{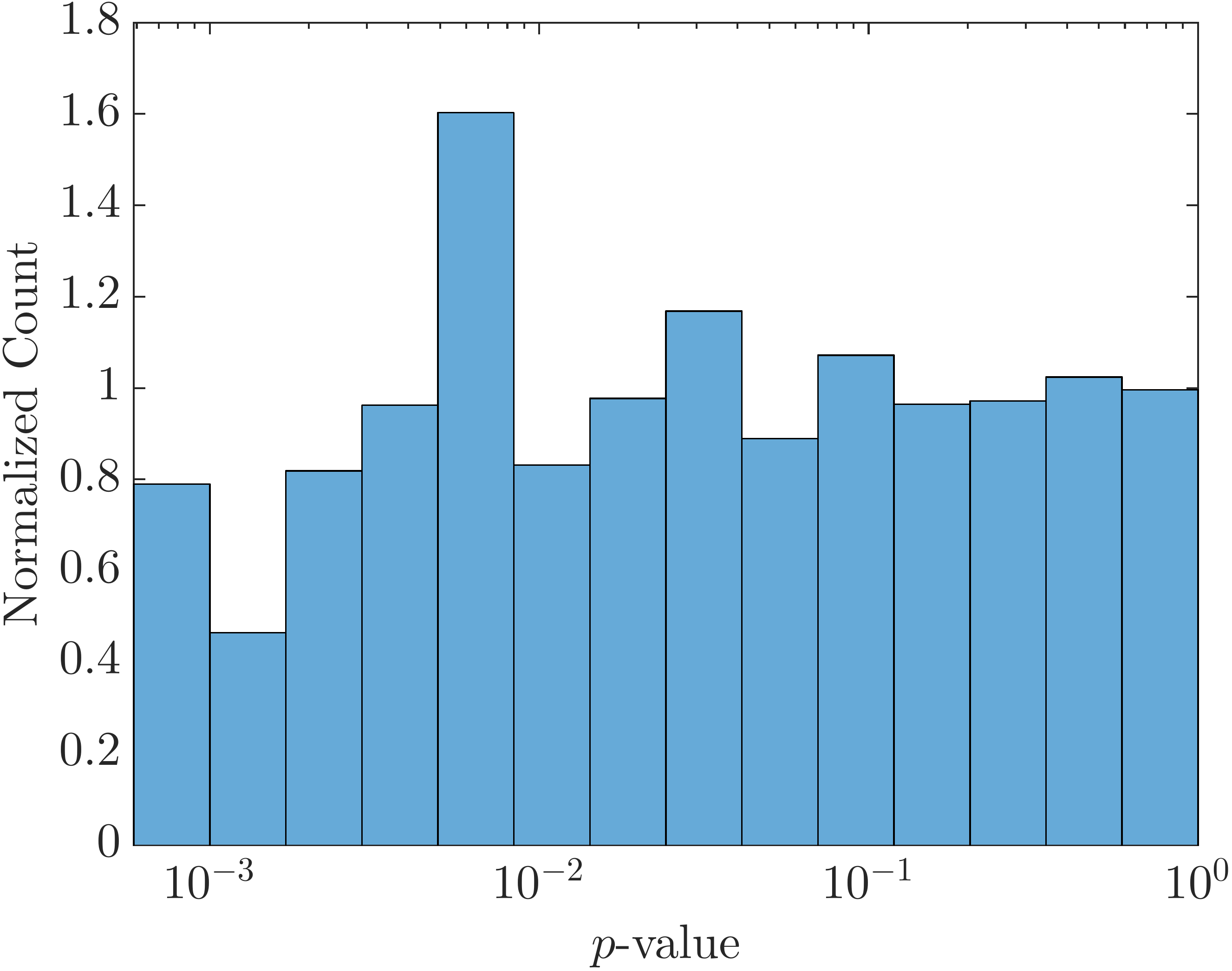}
		\label{subfig:pval_distribution_smica}
	}
	\subfigure[]{
		\includegraphics[width=0.47\textwidth]{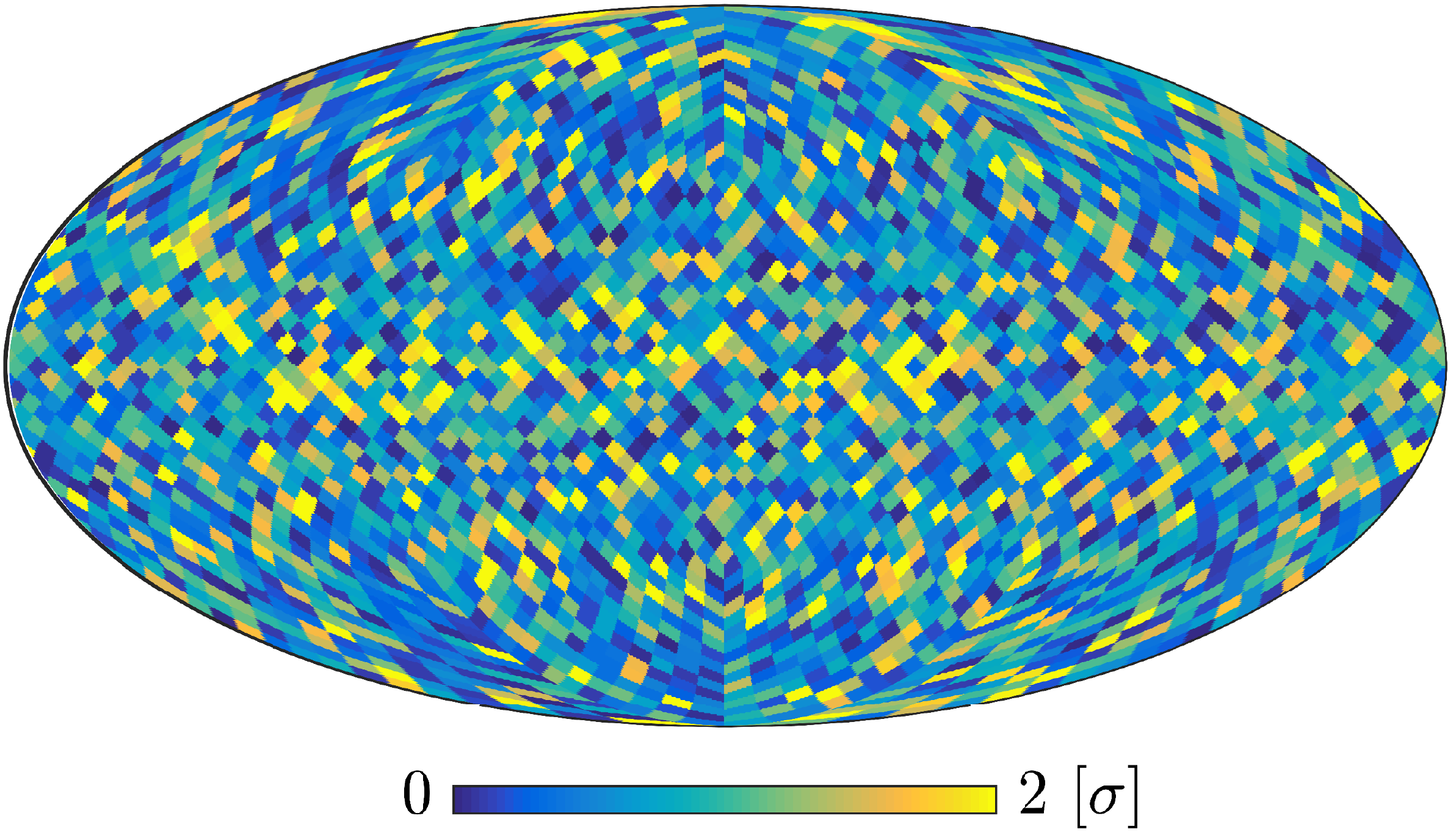}
		\label{subfig:sigma_map_smica}
	}
	\caption{\subref{subfig:pval_distribution_smica}~The distribution of $p$-values for the SMICA patches. The bin edges are logarithmically spaced and the histogram is normalized to unit total area. \subref{subfig:sigma_map_smica}~The $\sigma$-level for each patch of the SMICA map, in Galactic coordinates.\label{fig:Distribution_smica}}
\end{figure*}

The contour lines strongly suggest that the distribution of patch $p$-values for the SMICA map follows the expected distribution, with $\approx 0.1\%$ of the patches crossing the $3\sigma$ boundary. We use the mean histogram to assign a $p$-value to each patch as described above and plot in Fig.~\ref{subfig:pval_distribution_smica} the distribution of the values.

It is apparent that the distribution is quite uniform. This provides confirmation that the SMICA map is indeed consistent with our ensemble of realizations and provides further support for the conclusion of~\cite{Ben-David:2015sia} that the more complicated FFP simulations are not essential for analyzing the statistical properties of the SMICA map; simple Gaussian appear to be sufficient.  Moreover, in the context of the present analysis, this conclusion can be extended to smaller scales with $\lmax$ as large as $1000$.

Having labeled each patch with a $p$-value, or the corresponding $\sigma$-levels, we can now consider the spatial distribution of these values. We therefore plot in Fig.~\ref{subfig:sigma_map_smica} a map showing the $p$-values, translated to $\sigma$-levels, of the patches of the SMICA map.
Visual inspection of the SMICA patches suggests a tendency towards an excess of outliers in the Galactic plane, but we have not investigated this question quantitatively. 


\subsection{The Haslam Map} 
\label{sub:the_haslam_map}

In addition to the SMICA map of CMB temperature fluctuations, we also apply our method to a reprocessed version~\cite{Remazeilles:2014mba} of the 408~MHz radio map of Haslam et al.~\cite{Haslam:1982zz}. We adopt this map as typical of the maps used as templates for cleaning foreground contributions to the CMB data. Understanding the statistical properties of such maps, or at least finding outlying regions in them, could improve the accuracy of CMB extraction. Since it is primarily a map of Galactic emissions, we do not expect the Haslam map as a whole to be Gaussian-distributed. Working in small patches, however, allows us to inspect the distributions in local areas of the map that may or may not be Gaussian. The $56'$ resolution of the map enables us to choose a patch grid (with $\Nside=16$ but not smaller) identical to that chosen for the SMICA map. Since the Haslam map is provided in an $\Nside=512$ pixelization, we are again left with 1024 pixels in each patch.

\begin{figure*}
	\centering
	\subfigure[]{
		\includegraphics[width=0.47\textwidth]{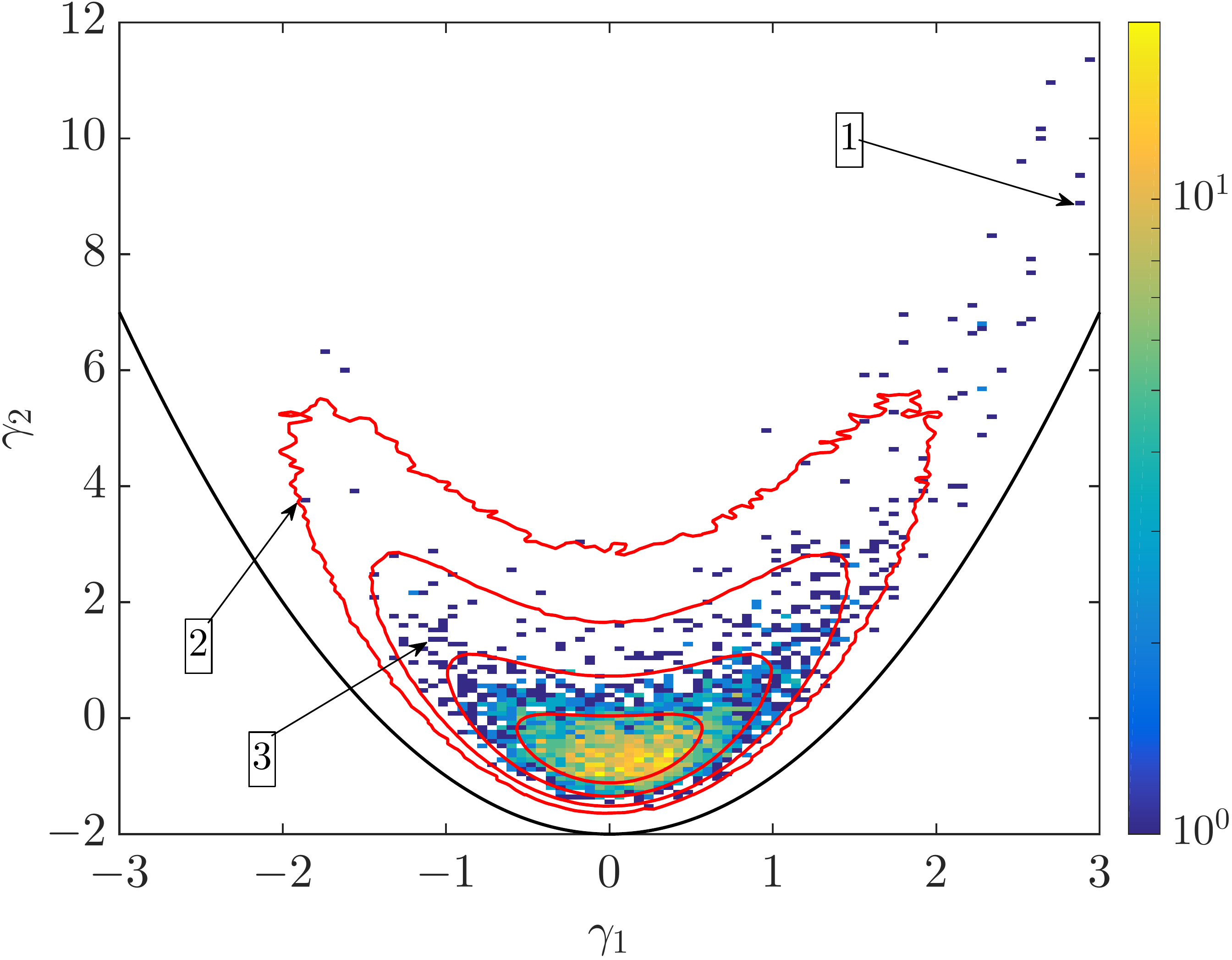}
		\label{subfig:histogram_haslam}
	}
	\subfigure[]{
		\includegraphics[width=0.47\textwidth]{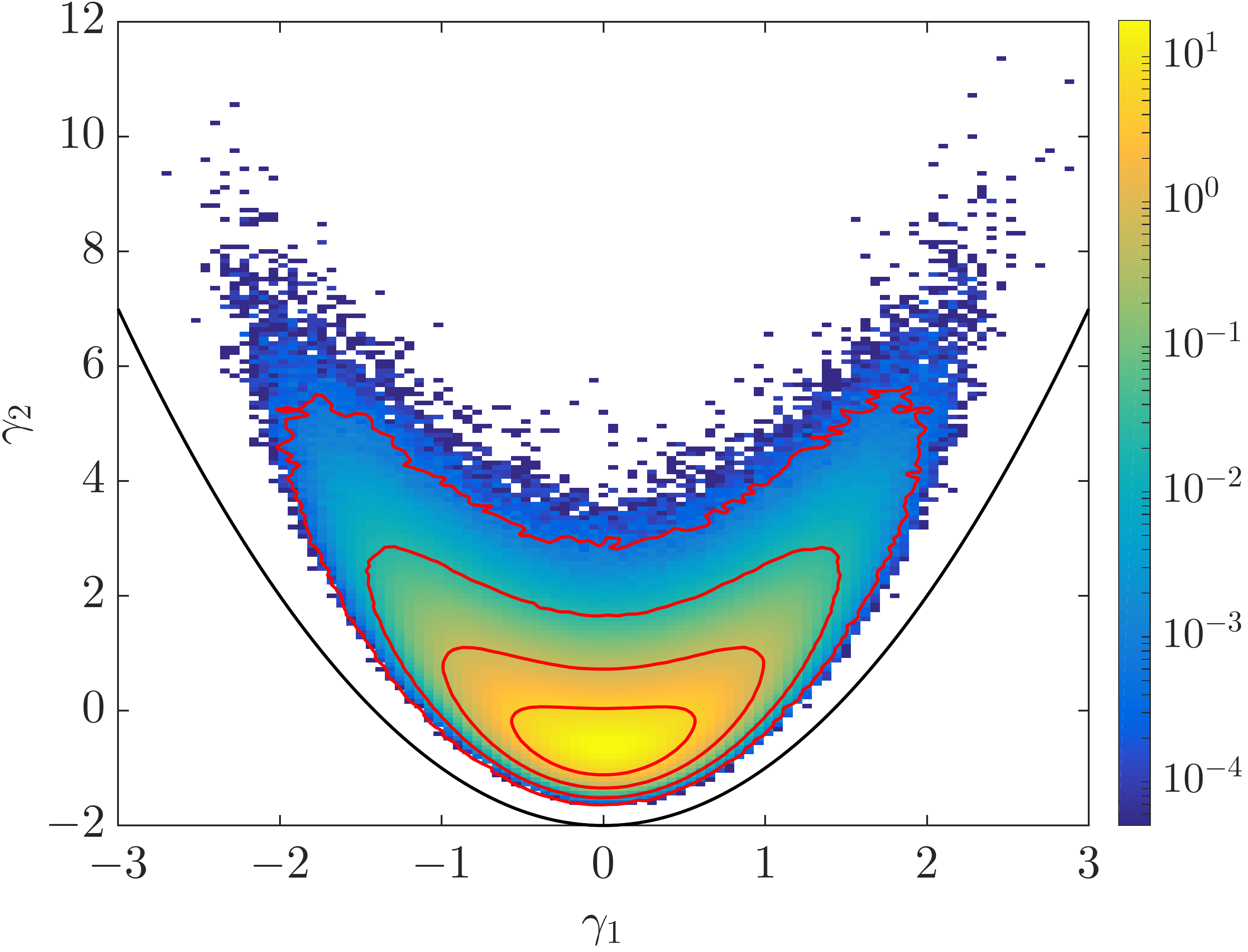}
		\label{subfig:mean_histogram_haslam}
	}
	\caption{\subref{subfig:histogram_haslam}~The skewness--kurtosis histogram for the Haslam map. The bins of the three patches given as examples below are marked with arrows. \subref{subfig:mean_histogram_haslam}~Mean histogram based on Gaussian realizations using the Haslam power spectrum, as explained in the text. The same set of 1--4$\sigma$ contour levels, calculated using the mean histogram in~\subref{subfig:mean_histogram_haslam}, is shown in both panels (\emph{red}). \label{fig:HaslamHistograms}}
\end{figure*}

The skewness--kurtosis histogram for the Haslam map is shown in Fig.~\ref{subfig:histogram_haslam}.
The range of values of $\gamma_1$ and $\gamma_2$  is far larger than that found in the case  of the SMICA map (Fig.~\ref{fig:histogram_smica}). It is also immediately apparent that it is much more probable for a Haslam patch to have positive skewness than negative. We will comment more on this asymmetry below.

As before, we require an ensemble in order to assign a $p$-value to each patch.  In contrast to the map of CMB fluctuations, however, the Haslam map is not believed to be a Gaussian realization drawn in harmonic space using some angular power spectrum.  Thus, it is not clear how to generate a suitable ensemble. 
We note first that the ensemble should be Gaussian. This is a result of \emph{our} wish to classify the patches by their level of Gaussianity and is unrelated to the intrinsic distribution of the data.  The determination of the nature of this Gaussian or, equivalently, of the corresponding correlation matrix of the Haslam map would require detailed knowledge of the specific physical processes that govern synchrotron emissions in the Galaxy. In the interests of simplicity and generality, we prefer using the correlation structure of the Haslam map itself as a reference point. In order to obtain a statistical measure of the correlations, we are then led to treat the map as a realization of a homogeneous and isotropic distribution. We transform the map to the spherical harmonic domain and calculate the observed power spectrum,
\be
\widehat{C}_\ell = \frac{1}{2\ell+1} \sum_{m=-\ell}^\ell |a_{\ell m}|^2,
\ee
where the $a_{\ell m}$ are the harmonic coefficients. We then use this power spectrum to draw Gaussian realizations in harmonic space.  This results in an ensemble of maps having, on average, the same structure of correlations as the Haslam map.  We take $\lmax=600$ since the Haslam angular power is negligible for smaller scales (larger values of $\ell$).  We draw $20\,000$ realizations, which allows us to reach a significance level of $4\sigma$.

\begin{figure*}
	\centering
	\subfigure[]{
		\includegraphics[width=0.47\textwidth]{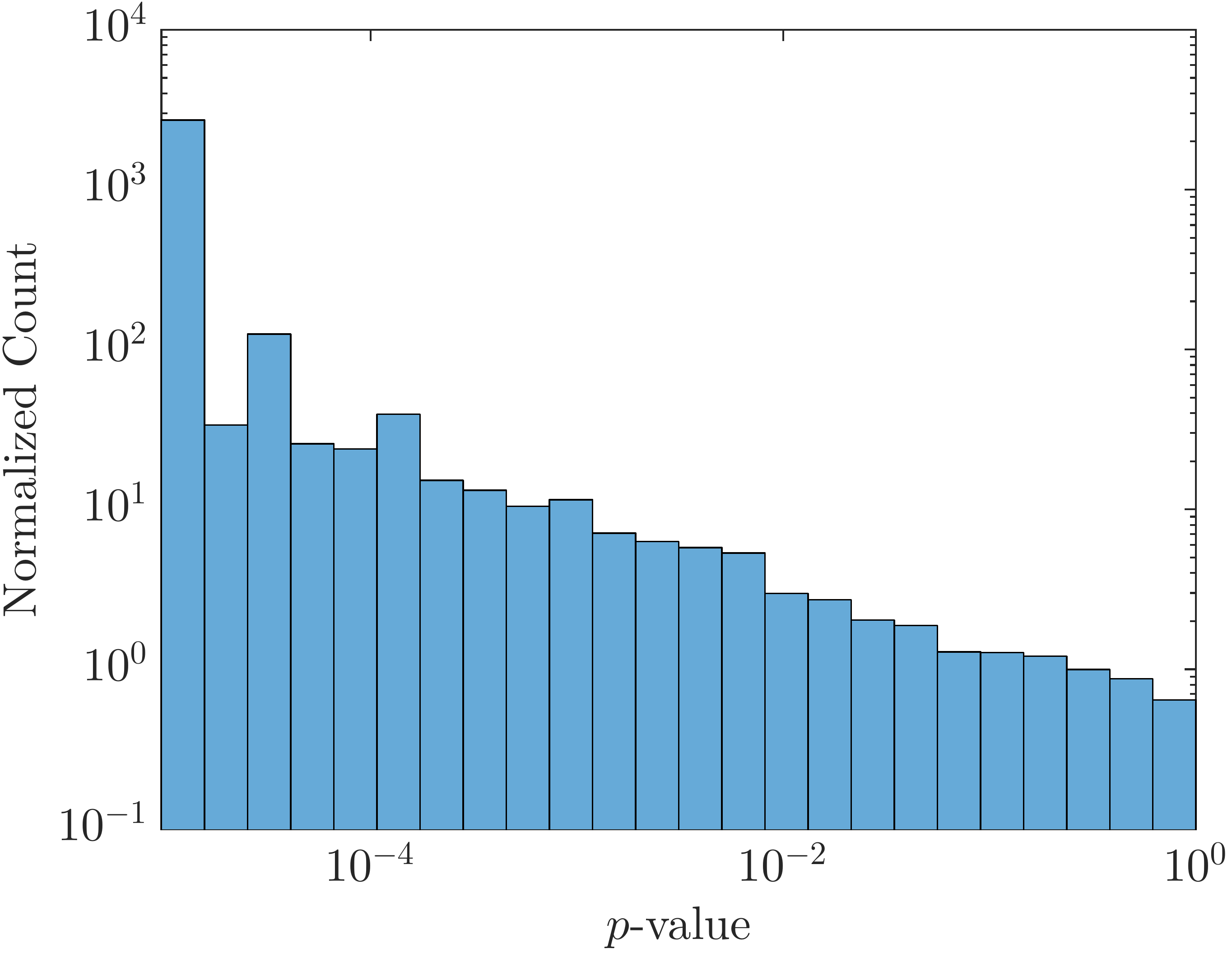}
		\label{subfig:pval_distribution_haslam}
	}
	\subfigure[]{
		\includegraphics[width=0.47\textwidth]{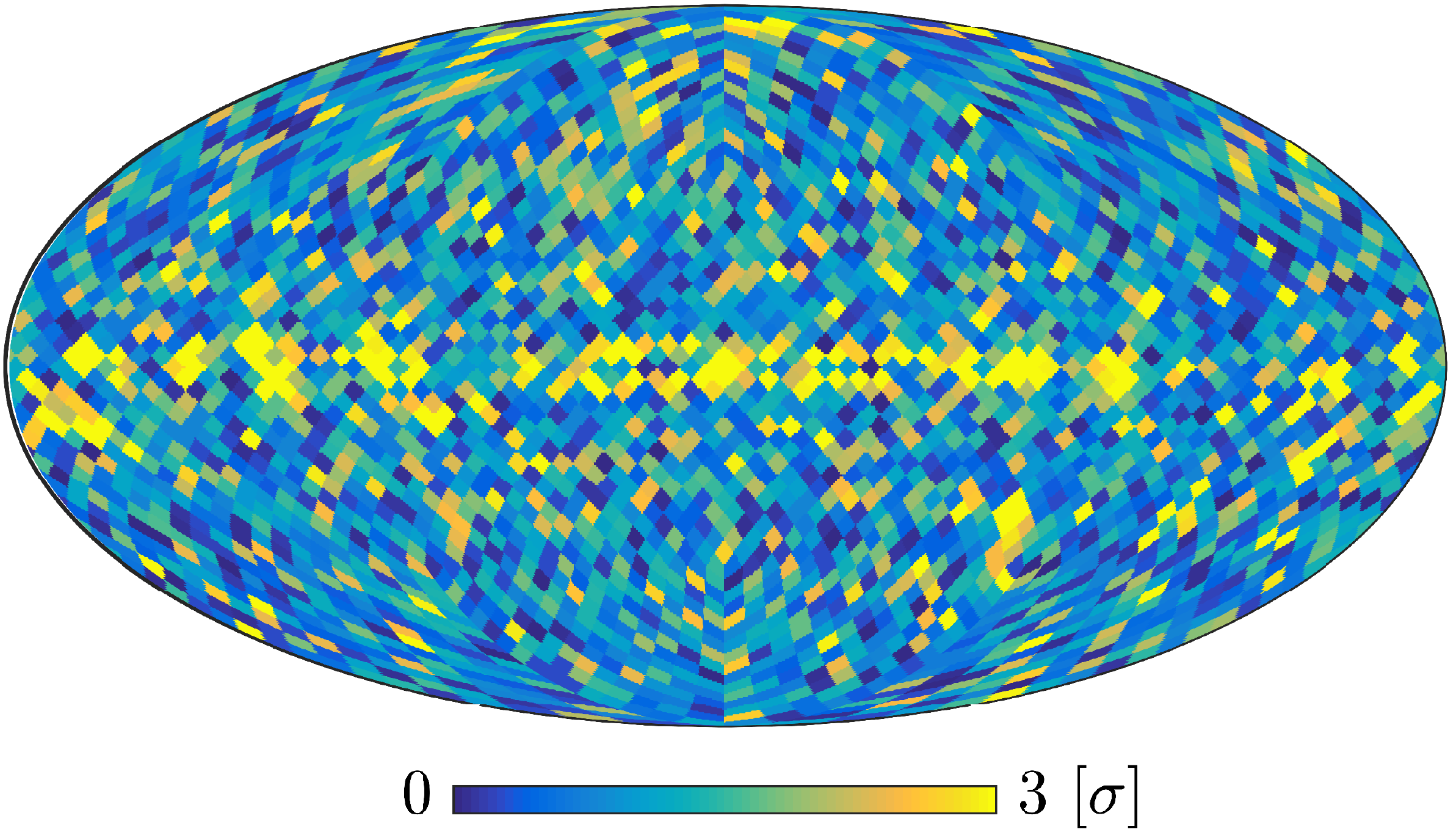}
		\label{subfig:sigma_map_haslam}
	}
	\caption{\subref{subfig:pval_distribution_haslam}~The distribution of $p$-values for the Haslam patches. The bin edges are logarithmically spaced and the histogram is normalized to unit total area.  We note that the leftmost column here represents the sum of all patches with $p \lesssim 10^{-5}$. \subref{subfig:sigma_map_haslam}~The $\sigma$-level for each patch of the Haslam map, in Galactic coordinates. \label{fig:Distribution_Haslam}}
\end{figure*}

The mean histogram obtained using this ensemble is shown in Fig.~\ref{subfig:mean_histogram_haslam}. We use it to calculate the 1--4$\sigma$ contour levels and superimpose them, as an illustration, on both histograms of Fig.~\ref{fig:HaslamHistograms}. It is clear that many patches of the Haslam map are located far beyond the $4\sigma$ contour and are therefore potentially interesting. However, it is also intriguing that so many of the patches are located within the $1\sigma$ contour in light of the fact that the map as a whole is highly non-Gaussian. This is an illustration of the advantage of an analysis of small patches; it can highlight local features while disregarding global features of the map. We can now assign $p$-values to the patches.   Their distribution is plotted in Fig.~\ref{subfig:pval_distribution_haslam}.
We see that the distribution is not uniform. This indicates, as expected, that the Haslam map as a whole is not a consistent realization of the ensemble.

As with the SMICA map, we can examine the spatial distribution of the patches labeled by their significance levels. This map is shown in Fig.~\ref{subfig:sigma_map_haslam}.
It is immediately apparent that many of the highly non-Gaussian patches are located in the area of the Galactic plane.  More surprisingly, it is evident that many of the patches in the Galactic plane are labeled with a very low value of $\sigma$.  (We will suggest below that this result does not necessarily indicate that these patches are free of non-Gaussian foreground effects.)  We also notice some highly non-Gaussian patches at high Galactic latitudes far from the Galactic plane.\\

The intriguing results of Fig.~\ref{subfig:sigma_map_haslam} provide motivation to examine some of the patches by eye in the hope of determining 
why they are anomalous. Our labeling of the patches by $p$-values (or $\sigma$-levels) provides a natural ordering scheme for their examination.
\begin{figure}
	\centering
	\includegraphics[width=\plotW]{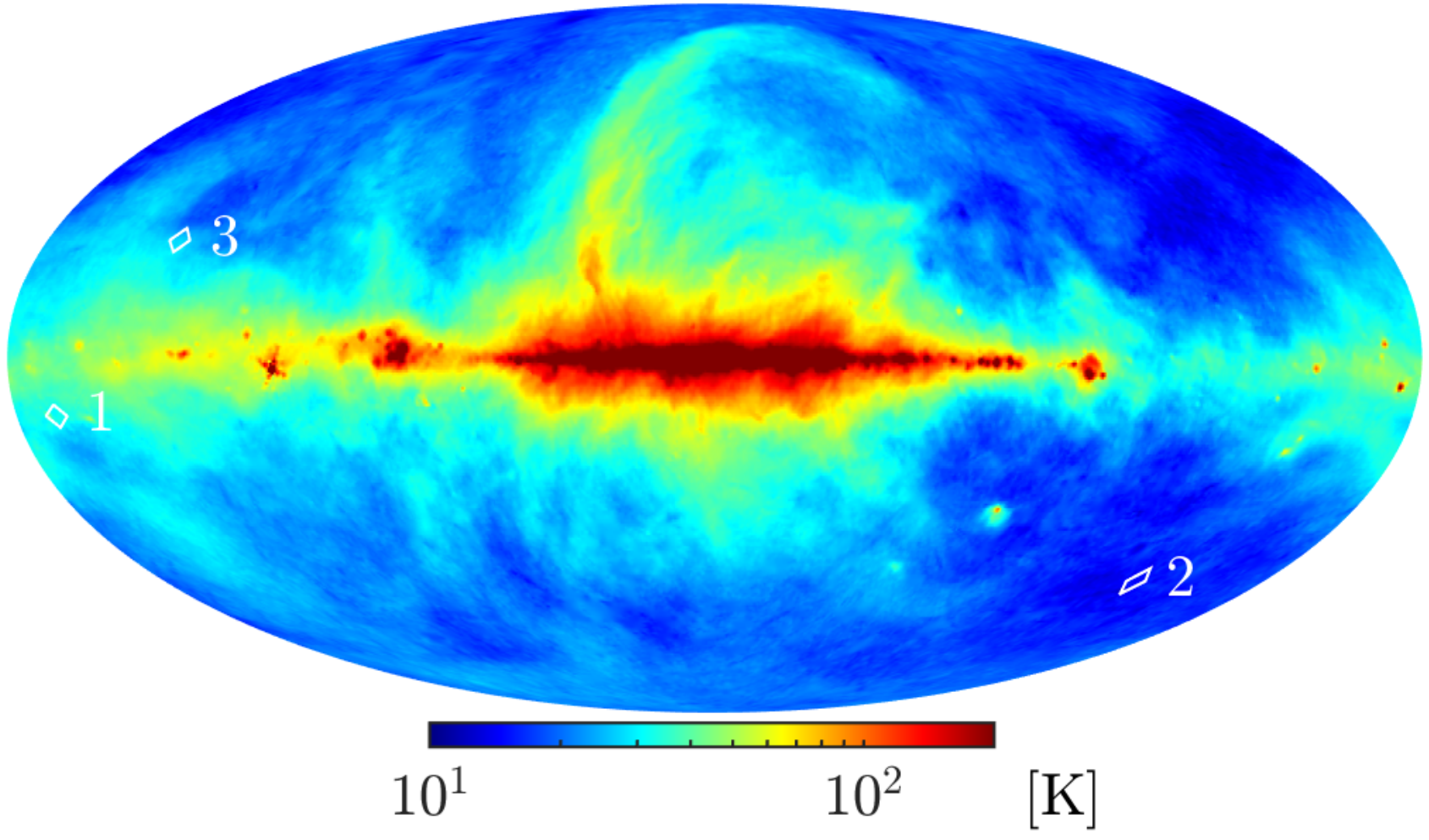}
	\caption{The reprocessed 408~MHz Haslam map, in Galactic coordinates. The locations of the three sample patches are marked in white. \label{fig:haslam_map}}
\end{figure}
We select three patches that are each characteristic of their locations in the skewness--kurtosis histogram, Fig.~\ref{subfig:histogram_haslam}. Their positions in this histogram are indicated by the labels `1', `2' and `3'.  In addition, in Fig.~\ref{fig:haslam_map} we show the Haslam map and indicate on it the location of the patches.  They lie in different areas of the map and have distinct characteristics.  The panels of Fig.~\ref{fig:patchescloseup} show the patches in question (enclosed in white) along with portions of their neighboring patches.  The first of these patches, Fig.~\ref{subfig:hotspot}, belongs to the relatively well-populated branch of the histogram with high kurtosis and large positive skewness. Such values arise naturally when the distribution of the patch contains a small area of relatively high intensity.  These conditions are easily produced by a point source as is the case in the figure shown.  The region of high intensity covers only a small fraction of the patch; the bulk has an intensity that is relatively low compared to the mean intensity of the patch.  Patches with similar distributions are most frequently found along the Galactic plane. This fact is clearly visible in Fig.~\ref{subfig:sigma_map_haslam} and incidentally provides qualitative confirmation of the success of the method in~\cite{Remazeilles:2014mba}, which was designed to remove extragalactic point sources from the original Haslam map.
\begin{figure*}
	\centering
	\subfigure[]{
		\includegraphics[height=0.24\textwidth]{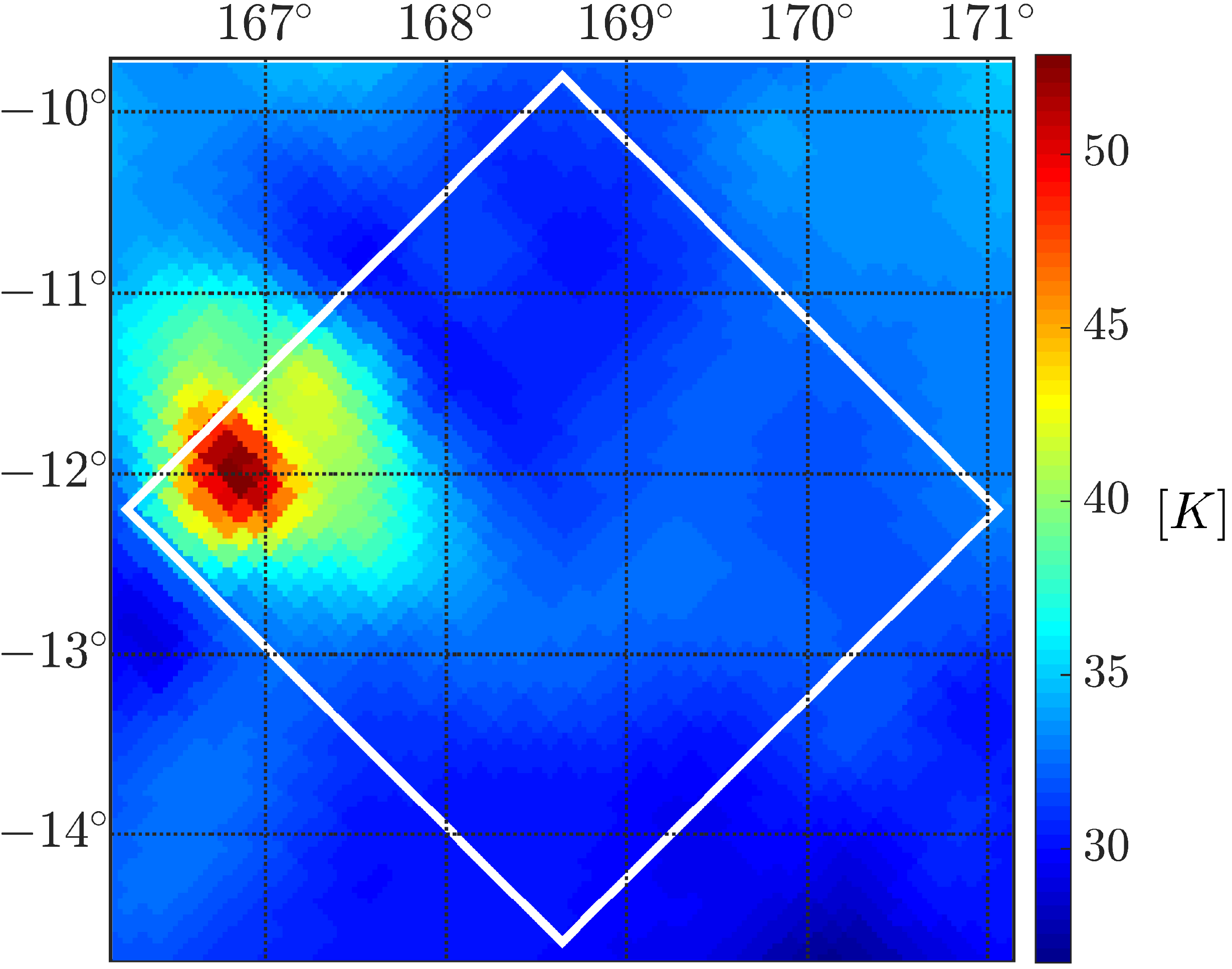}
		\label{subfig:hotspot}
	}
	\subfigure[]{
		\includegraphics[height=0.24\textwidth]{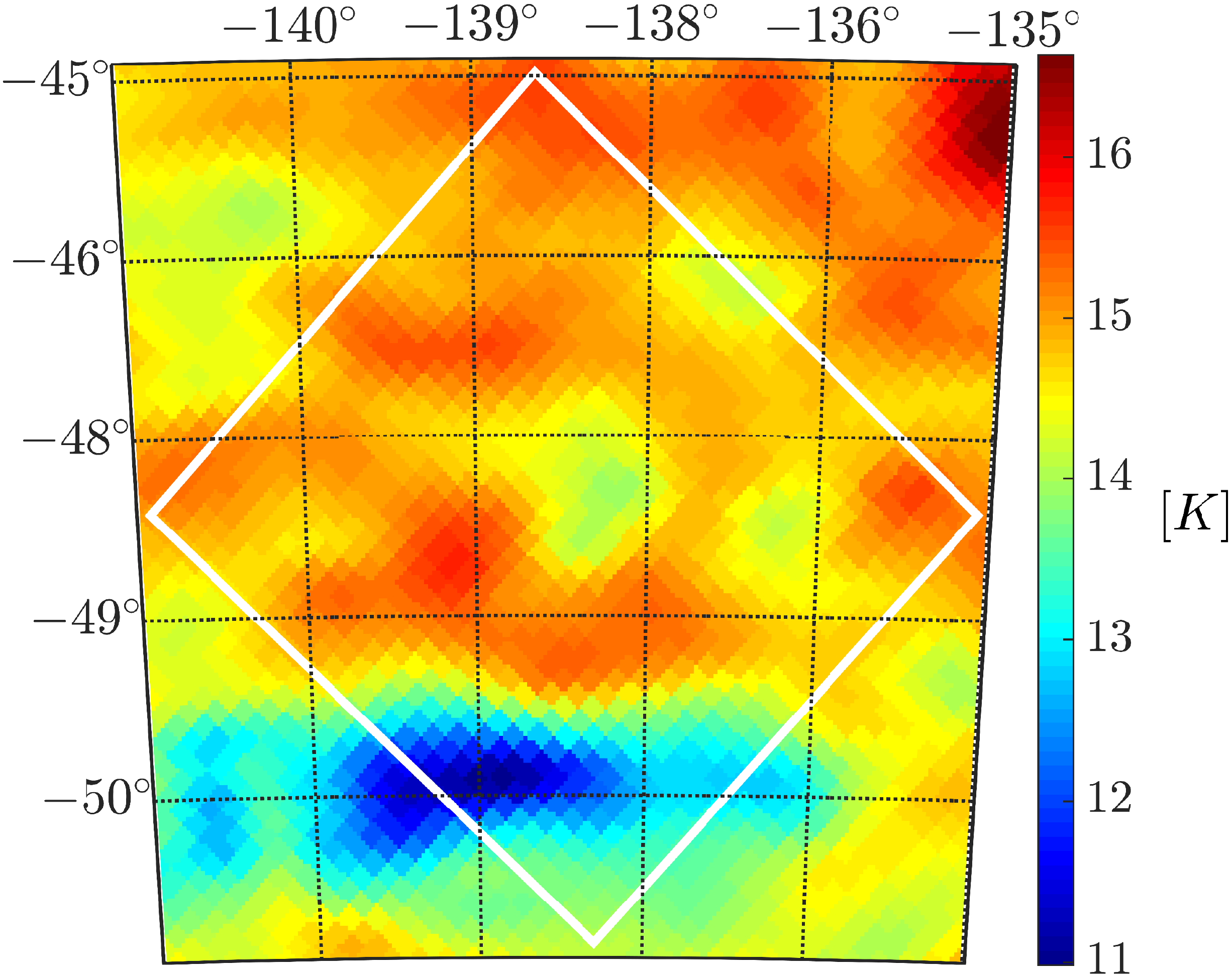}
		\label{subfig:coldspot}
	}
	\subfigure[]{
		\includegraphics[height=0.24\textwidth]{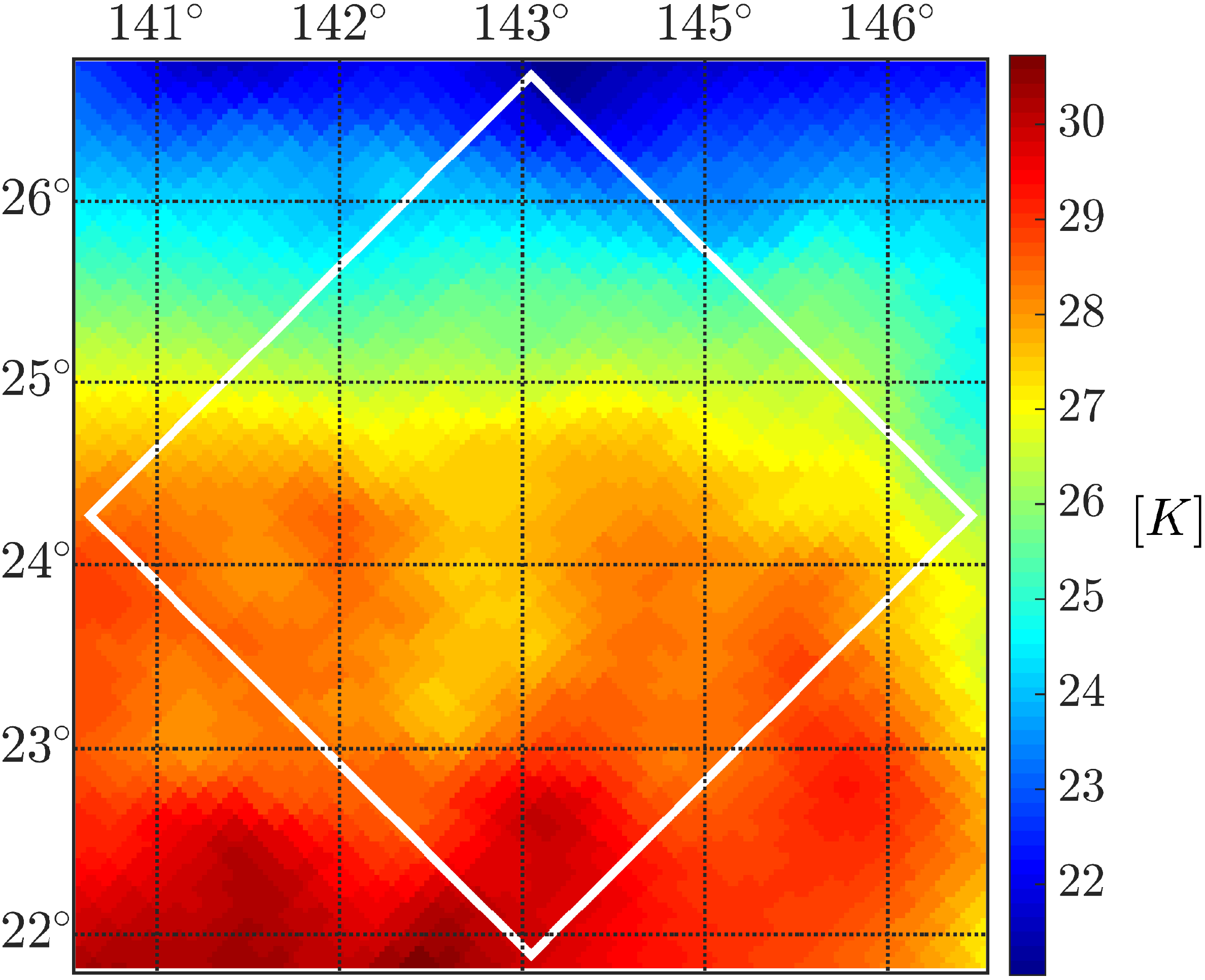}
		\label{subfig:gradual}
	}
	\caption{From left to right, patches 1, 2, and 3 and portions of their neighboring patches.  The vertical and horizontal axes show Galactic latitude and longitude, respectively.  The skewness and kurtosis of each patch can be seen in Fig.~\ref{subfig:histogram_haslam}, and Fig.~\ref{fig:haslam_map} shows their locations on the Haslam map. 	 \label{fig:patchescloseup}}
\end{figure*}

We now consider patch 2---a patch selected from the less populated branch of the histogram in Fig.~\ref{subfig:histogram_haslam}.  The large magnitudes of the kurtosis and skewness again ensure that the $p$-value of this patch is small.  In this case, however, the skewness is strongly negative.  The patch should 
contain a small region of very low intensity while the complementary area should have an intensity somewhat above the average value.  The obvious analogy to Fig.~\ref{subfig:hotspot} thus makes it reasonable to expect to find a cold spot or point sink.  Clearly, Fig.~\ref{subfig:coldspot} meets this expectation.  Although a cursory glance at the night sky is sufficient to remind us of the existence of point sources, it is more challenging to think of explanations for the phenomenon 
of cold spots.  There are several possible explanations for the existence of regions where the intensity of synchrotron radiation is low.  Small magnetic fields and/or low electron densities can reduce the intensity of radiation emitted from a given region.  Alternatively, intervening matter can absorb synchrotron radiation once it has been emitted.  These effects can work individually or in concert.  The relative paucity of patches with large negative skewness, evident in the histogram of Fig.~\ref{subfig:histogram_haslam}, suggests that localized radiation sinks are significantly less likely than bright point sources.

The results reported here were obtained from a single patching pattern.  It can happen, however, that the specific values of skewness and kurtosis of a given patch can be highly sensitive to its particular position on the map. Thus, a patch with high skewness and kurtosis need not necessarily contain a point source but can rather lie at an advantageous location, grazing a larger-scale structure.  This can also occur for other values $( \gamma_1 , \gamma_2)$, e.g., also simulating cold point sources.  Thus, patch 3 in Fig.~\ref{subfig:gradual} illustrates a case where the patch lies on top of a larger structure in which the temperature changes smoothly over the area of the patch.  The value of the skewness---including its sign---will depend on the precise location of the patch relative to this fixed large scale structure.  

The last case, not shown in a figure here, is that of random fluctuations. Large fractions of the sky are dominated by diffuse emission, which in the case of the Haslam map is closely connected to the morphology and fluctuations in the Galactic magnetic field and the local electron density. The pixels contained in patches lying in these regions will be distributed in a largely Gaussian manner.  The histogram, Fig.~\ref{subfig:histogram_haslam}, reveals that approximately 52\% of the patches in the Haslam map are Gaussian with $\sigma \le 1$, and inspection of Fig.~\ref{subfig:sigma_map_haslam} provides visual confirmation.  The dominance of Gaussian patches was expected for the SMICA map and is quantitatively confirmed by Figs.~\ref{fig:histogram_smica}--\ref{fig:Distribution_smica}.  A similar dominance in the Haslam map was {\em not\/} expected.  We will return to this surprising result in the following section.  

\section{Discussion} 
\label{sec:discussion}

The results presented here are intended to illustrate the role that higher moments can play in understanding the statistical behaviour of foreground maps.  In practice, it  is necessary to adjust the size of the patches to ensure that two obvious criteria are met:  The patches must be large on a scale set by the correlation angle, $\theta_{\rm c}$, in order to ensure that individual pixels have a  reasonable degree of statistical independence.  Second, the patches must be small in comparison with the size of diffuse foreground effects.  Such considerations led us to choose patches with $N_{\rm side} = 16$.  Clearly, patch size can be tuned to the scale of phenomena that we would like to identify.  For example, the southern hemisphere cold spot seen in both WMAP and CMB temperature maps has an angular size of approximately $5^{\circ}$.  As a consequence, it does not appear to be exceptional in the SMICA plot of Fig.~\ref{subfig:sigma_map_smica} that was obtained with a patch size of $3.7^{\circ}$.  It should be easy to spot with a larger patch size.  We have also restricted our attention to a single realization of the ``tiling'' of the full sky.  Since we have noted some phenomena that can be sensitive to the precise location of a patch, it would be prudent to consider several tilings displaced by an amount comparable to the patch size.  Patch selection here was made using the \textsc{HEALP}ix package that produces a subdivision of a spherical surface into patches of equal area, but our method can evidently be used for other patch forms and sizes.  We also note that, although the decision to consider the statistics of skewness and kurtosis was natural, the same approach could be expanded to include a larger number of higher moments.  The only technical challenge in such an extension would be the determination of the inequalities analogous to Eq.~\eqref{eq:pearson} that such higher moments must satisfy.\\

Turning to the results of our analysis, we note that the SMICA map of CMB temperature fluctuations is entirely consistent with the expectation of 
Gaussian behaviour.  This agreement is confirmed with considerable accuracy by the fact that the distribution of $p$-values shown in Fig.~\ref{subfig:pval_distribution_smica} is independent of $p$.\footnote{Bear in mind, for example, that the number of patches contributing to this histogram 
decreases rapidly as $p \to 0$.}

With a far larger fraction of patches with small $p$-values, the Haslam map would appear to confirm the a priori belief that this foreground map is strongly non-Gaussian.  Closer inspection suggests that this conclusion may be premature.   Figs.~\ref{subfig:hotspot}--\ref{subfig:gradual} show that classification of a patch by its $p$-value alone is not sufficient to describe its full physical content.   These figures reveal three classes of anomalous patches that can be distinguished by their structure in pixel space: point sources, cold spots and strong intensity gradients.  Although it was not our goal to identify such structures, our method has the potential to do so mechanically.  In practice, implementation should include shifting the location and changing the size of patches in order to avoid misidentification.  
Such localized phenomena differ distinctly from the diffuse effects of synchrotron radiation and merit special treatment when sky maps are to be cleaned.  Their removal from Fig.~\ref{subfig:histogram_haslam} would eliminate many of the patches with $\sigma > 3$ and virtually all patches with $\sigma > 4$.  The resulting histogram would have a significantly greater resemblance to a Gaussian result such as that shown in Fig.~\ref{subfig:mean_histogram_haslam}.  There are two distinct explanations for this somewhat unexpected result.  First, we note that Figs.~\ref{subfig:sigma_map_smica} and \ref{subfig:sigma_map_haslam} contain a large number of patches with $p \approx 1$ in the Galactic plane in spite of the fact that this region should be filled with sources of foreground effects.  Mertsch and Sarkar~\cite{Mertsch:2013pua} have argued that many independent foreground contributions can combine to create Gaussian distributions as a consequence of the central limit theorem.\footnote{While this may not be the explanation of the global abundance of Gaussian patches seen in the Haslam map, it provides a useful reminder that the absence of evidence for the existence of non-Gaussian foreground effects does {\em not\/} constitute evidence of their absence.}  For high Galactic latitudes, it is probably of greater importance to reconsider the assumptions of homogeneity and isotropy that led to the Gaussian realizations of the Haslam map shown in Fig.~\ref{subfig:mean_histogram_haslam}.  While these assumptions are not expected to be valid for the map as a whole, they may well be valid locally on the relatively small scale of the patch size.  In other words, large scale diffuse foreground effects can be described by correlations between spherical harmonics of relatively 
low order, $\ell \le \ell_0$ without the involvement of components with $\ell > \ell_0$.  These correlated terms will be approximately constant over patches whose characteristic size is smaller than $1/\ell_0$.  They will not contribute to dimensionless moments such as the skewness and kurtosis that will be determined by the  uncorrelated harmonics with large $\ell$.  In either language, no deeper explanation is required to understand the abundance of Gaussian patches.   

It is of some interest to consider the consequences of the present results for a single portion of the sky.  We have looked at the patches of the Haslam map that are associated with the region explored by the BICEP2 collaboration.  We have sorted these patches according to their significance in six bins of width $\sigma/2$ ranging from $0$ to $3 \sigma$.  Since the edges of the BICEP2 zone are somewhat loosely defined, we have considered both a larger and a smaller zone.  In the former case we considered all $58$ patches that touch the BICEP2 zone and found the distribution $[14,17,13,6,5,3]$ which can be compared with the Gaussian expectation of $[22.2,17.4,10.6,5.1,1.9,0.5]$.  The small zone consists of those $20$ patches that lie completely within the BICEP2 zone. In this case the distribution is $[8,6,4,2,0,0]$ which can be compared with the Gaussian result of $[7.7,6.0,3.7,1.8,0.7,0]$.  In both cases, there is good agreement with the Gaussian expectations.  Since the coupling between the foreground effects of synchrotron radiation and dust emission are expected to be large, there would appear to be grounds for the optimistic hope that the dominate foreground effects in the BICEP2 zone are actually Gaussian.

CMB science is now entering an era of high precision polarization analysis that will require a more precise foreground separation than was needed for the temperature analysis.  We believe that the results presented here demonstrate that a patch-by-patch statistical study of the scale-free higher moments of the  
one-point distribution function can make a significant contribution to this program.


\begin{acknowledgments}
This work is based on observations obtained with Planck, an ESA science mission with instruments and contributions directly funded by ESA Member States, NASA, and Canada.  We gratefully acknowledge the financial support of Danmarks Grundforskningsfond and of the Villum Foundation.  It is a particular pleasure to thank Pavel Naselsky for the benefit of his invaluable advice and for his enthusiastic support.
\end{acknowledgments}

\bibliography{LocalSkewKurt}

\begin{thebibliography}{18}%
\makeatletter
\providecommand \@ifxundefined [1]{%
 \@ifx{#1\undefined}
}%
\providecommand \@ifnum [1]{%
 \ifnum #1\expandafter \@firstoftwo
 \else \expandafter \@secondoftwo
 \fi
}%
\providecommand \@ifx [1]{%
 \ifx #1\expandafter \@firstoftwo
 \else \expandafter \@secondoftwo
 \fi
}%
\providecommand \natexlab [1]{#1}%
\providecommand \enquote  [1]{``#1''}%
\providecommand \bibnamefont  [1]{#1}%
\providecommand \bibfnamefont [1]{#1}%
\providecommand \citenamefont [1]{#1}%
\providecommand \href@noop [0]{\@secondoftwo}%
\providecommand \href [0]{\begingroup \@sanitize@url \@href}%
\providecommand \@href[1]{\@@startlink{#1}\@@href}%
\providecommand \@@href[1]{\endgroup#1\@@endlink}%
\providecommand \@sanitize@url [0]{\catcode `\\12\catcode `\$12\catcode
  `\&12\catcode `\#12\catcode `\^12\catcode `\_12\catcode `\%12\relax}%
\providecommand \@@startlink[1]{}%
\providecommand \@@endlink[0]{}%
\providecommand \url  [0]{\begingroup\@sanitize@url \@url }%
\providecommand \@url [1]{\endgroup\@href {#1}{\urlprefix }}%
\providecommand \urlprefix  [0]{URL }%
\providecommand \Eprint [0]{\href }%
\providecommand \doibase [0]{http://dx.doi.org/}%
\providecommand \selectlanguage [0]{\@gobble}%
\providecommand \bibinfo  [0]{\@secondoftwo}%
\providecommand \bibfield  [0]{\@secondoftwo}%
\providecommand \translation [1]{[#1]}%
\providecommand \BibitemOpen [0]{}%
\providecommand \bibitemStop [0]{}%
\providecommand \bibitemNoStop [0]{.\EOS\space}%
\providecommand \EOS [0]{\spacefactor3000\relax}%
\providecommand \BibitemShut  [1]{\csname bibitem#1\endcsname}%
\let\auto@bib@innerbib\@empty
\bibitem [{\citenamefont {Adam}\ \emph
  {et~al.}(2015{\natexlab{a}})\citenamefont {Adam} \emph
  {et~al.}}]{Adam:2015rua}%
  \BibitemOpen
  \bibfield  {author} {\bibinfo {author} {\bibfnamefont {R.}~\bibnamefont
  {Adam}} \emph {et~al.} (\bibinfo {collaboration} {Planck Collaboration}),\
  }\href@noop {} {\enquote {\bibinfo {title} {{Planck 2015 results. I. Overview
  of products and scientific results}},}\ } (\bibinfo {year}
  {2015}{\natexlab{a}}),\ \bibinfo {note} {submitted to A\&A},\ \Eprint
  {http://arxiv.org/abs/1502.01582} {arXiv:1502.01582 [astro-ph.CO]}
  \BibitemShut {NoStop}%
\bibitem [{\citenamefont {Ade}\ \emph {et~al.}(2015{\natexlab{a}})\citenamefont
  {Ade} \emph {et~al.}}]{Ade:2015xua}%
  \BibitemOpen
  \bibfield  {author} {\bibinfo {author} {\bibfnamefont {P.~A.~R.}\
  \bibnamefont {Ade}} \emph {et~al.} (\bibinfo {collaboration} {Planck
  Collaboration}),\ }\href@noop {} {\enquote {\bibinfo {title} {{Planck 2015
  results. XIII. Cosmological parameters}},}\ } (\bibinfo {year}
  {2015}{\natexlab{a}}),\ \bibinfo {note} {submitted to A\&A},\ \Eprint
  {http://arxiv.org/abs/1502.01589} {arXiv:1502.01589 [astro-ph.CO]}
  \BibitemShut {NoStop}%
\bibitem [{\citenamefont {Mertsch}\ and\ \citenamefont
  {Sarkar}(2013)}]{Mertsch:2013pua}%
  \BibitemOpen
  \bibfield  {author} {\bibinfo {author} {\bibfnamefont {P.}~\bibnamefont
  {Mertsch}}\ and\ \bibinfo {author} {\bibfnamefont {S.}~\bibnamefont
  {Sarkar}},\ }\href {\doibase 10.1088/1475-7516/2013/06/041} {\bibfield
  {journal} {\bibinfo  {journal} {JCAP}\ }\textbf {\bibinfo {volume} {1306}},\
  \bibinfo {pages} {041} (\bibinfo {year} {2013})},\ \Eprint
  {http://arxiv.org/abs/1304.1078} {arXiv:1304.1078 [astro-ph.GA]} \BibitemShut
  {NoStop}%
\bibitem [{\citenamefont {Regis}(2011)}]{Regis:2011ji}%
  \BibitemOpen
  \bibfield  {author} {\bibinfo {author} {\bibfnamefont {M.}~\bibnamefont
  {Regis}},\ }\href {\doibase 10.1016/j.astropartphys.2011.07.003} {\bibfield
  {journal} {\bibinfo  {journal} {Astropart. Phys.}\ }\textbf {\bibinfo
  {volume} {35}},\ \bibinfo {pages} {170} (\bibinfo {year} {2011})},\ \Eprint
  {http://arxiv.org/abs/1101.5524} {arXiv:1101.5524 [astro-ph.HE]} \BibitemShut
  {NoStop}%
\bibitem [{\citenamefont {Kamionkowski}\ and\ \citenamefont
  {Kovetz}(2014)}]{PhysRevLett.113.191303}%
  \BibitemOpen
  \bibfield  {author} {\bibinfo {author} {\bibfnamefont {M.}~\bibnamefont
  {Kamionkowski}}\ and\ \bibinfo {author} {\bibfnamefont {E.~D.}\ \bibnamefont
  {Kovetz}},\ }\href {\doibase 10.1103/PhysRevLett.113.191303} {\bibfield
  {journal} {\bibinfo  {journal} {Phys. Rev. Lett.}\ }\textbf {\bibinfo
  {volume} {113}},\ \bibinfo {pages} {191303} (\bibinfo {year}
  {2014})}\BibitemShut {NoStop}%
\bibitem [{\citenamefont {Ade}\ \emph {et~al.}(2015{\natexlab{b}})\citenamefont
  {Ade} \emph {et~al.}}]{Ade:2015hxq}%
  \BibitemOpen
  \bibfield  {author} {\bibinfo {author} {\bibfnamefont {P.~A.~R.}\
  \bibnamefont {Ade}} \emph {et~al.} (\bibinfo {collaboration} {Planck
  Collaboration}),\ }\href@noop {} {\enquote {\bibinfo {title} {{Planck 2015
  results. XVI. Isotropy and statistics of the CMB}},}\ } (\bibinfo {year}
  {2015}{\natexlab{b}}),\ \bibinfo {note} {submitted to A\&A},\ \Eprint
  {http://arxiv.org/abs/1506.07135} {arXiv:1506.07135 [astro-ph.CO]}
  \BibitemShut {NoStop}%
\bibitem [{\citenamefont {Ben-David}\ \emph {et~al.}(2015)\citenamefont
  {Ben-David}, \citenamefont {Liu},\ and\ \citenamefont
  {Jackson}}]{Ben-David:2015sia}%
  \BibitemOpen
  \bibfield  {author} {\bibinfo {author} {\bibfnamefont {A.}~\bibnamefont
  {Ben-David}}, \bibinfo {author} {\bibfnamefont {H.}~\bibnamefont {Liu}}, \
  and\ \bibinfo {author} {\bibfnamefont {A.~D.}\ \bibnamefont {Jackson}},\
  }\href {\doibase 10.1088/1475-7516/2015/06/051} {\bibfield  {journal}
  {\bibinfo  {journal} {JCAP}\ }\textbf {\bibinfo {volume} {1506}},\ \bibinfo
  {pages} {051} (\bibinfo {year} {2015})},\ \Eprint
  {http://arxiv.org/abs/1506.07724} {arXiv:1506.07724 [astro-ph.CO]}
  \BibitemShut {NoStop}%
\bibitem [{\citenamefont {Haslam}\ \emph {et~al.}(1982)\citenamefont {Haslam},
  \citenamefont {Salter}, \citenamefont {Stoffel},\ and\ \citenamefont
  {Wilson}}]{Haslam:1982zz}%
  \BibitemOpen
  \bibfield  {author} {\bibinfo {author} {\bibfnamefont {C.~G.~T.}\
  \bibnamefont {Haslam}}, \bibinfo {author} {\bibfnamefont {C.~J.}\
  \bibnamefont {Salter}}, \bibinfo {author} {\bibfnamefont {H.}~\bibnamefont
  {Stoffel}}, \ and\ \bibinfo {author} {\bibfnamefont {W.~E.}\ \bibnamefont
  {Wilson}},\ }\href@noop {} {\bibfield  {journal} {\bibinfo  {journal}
  {Astron. Astrophys. Suppl. Ser.}\ }\textbf {\bibinfo {volume} {47}},\
  \bibinfo {pages} {1} (\bibinfo {year} {1982})}\BibitemShut {NoStop}%
\bibitem [{\citenamefont {Remazeilles}\ \emph {et~al.}(2015)\citenamefont
  {Remazeilles}, \citenamefont {Dickinson}, \citenamefont {Banday},
  \citenamefont {Bigot-Sazy},\ and\ \citenamefont
  {Ghosh}}]{Remazeilles:2014mba}%
  \BibitemOpen
  \bibfield  {author} {\bibinfo {author} {\bibfnamefont {M.}~\bibnamefont
  {Remazeilles}}, \bibinfo {author} {\bibfnamefont {C.}~\bibnamefont
  {Dickinson}}, \bibinfo {author} {\bibfnamefont {A.~J.}\ \bibnamefont
  {Banday}}, \bibinfo {author} {\bibfnamefont {M.~A.}\ \bibnamefont
  {Bigot-Sazy}}, \ and\ \bibinfo {author} {\bibfnamefont {T.}~\bibnamefont
  {Ghosh}},\ }\href {\doibase 10.1093/mnras/stv1274} {\bibfield  {journal}
  {\bibinfo  {journal} {Mon. Not. R. Astron. Soc.}\ }\textbf {\bibinfo {volume}
  {451}},\ \bibinfo {pages} {4311} (\bibinfo {year} {2015})},\ \Eprint
  {http://arxiv.org/abs/1411.3628} {arXiv:1411.3628 [astro-ph.IM]} \BibitemShut
  {NoStop}%
\bibitem [{\citenamefont {Adam}\ \emph
  {et~al.}(2015{\natexlab{b}})\citenamefont {Adam} \emph
  {et~al.}}]{Adam:2015wua}%
  \BibitemOpen
  \bibfield  {author} {\bibinfo {author} {\bibfnamefont {R.}~\bibnamefont
  {Adam}} \emph {et~al.} (\bibinfo {collaboration} {Planck Collaboration}),\
  }\href@noop {} {\enquote {\bibinfo {title} {{Planck 2015 results. X. Diffuse
  component separation: Foreground maps}},}\ } (\bibinfo {year}
  {2015}{\natexlab{b}}),\ \bibinfo {note} {submitted to A\&A},\ \Eprint
  {http://arxiv.org/abs/1502.01588} {arXiv:1502.01588 [astro-ph.CO]}
  \BibitemShut {NoStop}%
\bibitem [{\citenamefont {Adam}\ \emph
  {et~al.}(2015{\natexlab{c}})\citenamefont {Adam} \emph
  {et~al.}}]{Adam:2015tpy}%
  \BibitemOpen
  \bibfield  {author} {\bibinfo {author} {\bibfnamefont {R.}~\bibnamefont
  {Adam}} \emph {et~al.} (\bibinfo {collaboration} {Planck Collaboration}),\
  }\href@noop {} {\enquote {\bibinfo {title} {{Planck 2015 results. IX. Diffuse
  component separation: CMB maps}},}\ } (\bibinfo {year}
  {2015}{\natexlab{c}}),\ \bibinfo {note} {submitted to A\&A},\ \Eprint
  {http://arxiv.org/abs/1502.05956} {arXiv:1502.05956 [astro-ph.CO]}
  \BibitemShut {NoStop}%
\bibitem [{\citenamefont {Pearson}(1916)}]{Pearson429}%
  \BibitemOpen
  \bibfield  {author} {\bibinfo {author} {\bibfnamefont {K.}~\bibnamefont
  {Pearson}},\ }\href {\doibase 10.1098/rsta.1916.0009} {\bibfield  {journal}
  {\bibinfo  {journal} {Phil. Trans. R. Soc. A}\ }\textbf {\bibinfo {volume}
  {216}},\ \bibinfo {pages} {429} (\bibinfo {year} {1916})}\BibitemShut
  {NoStop}%
\bibitem [{\citenamefont {Rohatgi}\ and\ \citenamefont
  {Székely}(1989)}]{Rohatgi1989297}%
  \BibitemOpen
  \bibfield  {author} {\bibinfo {author} {\bibfnamefont {V.~K.}\ \bibnamefont
  {Rohatgi}}\ and\ \bibinfo {author} {\bibfnamefont {G.~J.}\ \bibnamefont
  {Székely}},\ }\href {\doibase 10.1016/0167-7152(89)90035-7} {\bibfield
  {journal} {\bibinfo  {journal} {Statistics \& Probability Letters}\ }\textbf
  {\bibinfo {volume} {8}},\ \bibinfo {pages} {297 } (\bibinfo {year}
  {1989})}\BibitemShut {NoStop}%
\bibitem [{\citenamefont {Klaassen}\ \emph {et~al.}(2000)\citenamefont
  {Klaassen}, \citenamefont {Mokveld},\ and\ \citenamefont {van
  Es}}]{Klaassen2000131}%
  \BibitemOpen
  \bibfield  {author} {\bibinfo {author} {\bibfnamefont {C.~A.}\ \bibnamefont
  {Klaassen}}, \bibinfo {author} {\bibfnamefont {P.~J.}\ \bibnamefont
  {Mokveld}}, \ and\ \bibinfo {author} {\bibfnamefont {B.}~\bibnamefont {van
  Es}},\ }\href {\doibase 10.1016/S0167-7152(00)00090-0} {\bibfield  {journal}
  {\bibinfo  {journal} {Statistics \& Probability Letters}\ }\textbf {\bibinfo
  {volume} {50}},\ \bibinfo {pages} {131 } (\bibinfo {year}
  {2000})}\BibitemShut {NoStop}%
\bibitem [{\citenamefont {Jarque}\ and\ \citenamefont
  {Bera}(1980)}]{Jarque1980255}%
  \BibitemOpen
  \bibfield  {author} {\bibinfo {author} {\bibfnamefont {C.~M.}\ \bibnamefont
  {Jarque}}\ and\ \bibinfo {author} {\bibfnamefont {A.~K.}\ \bibnamefont
  {Bera}},\ }\href {\doibase 10.1016/0165-1765(80)90024-5} {\bibfield
  {journal} {\bibinfo  {journal} {Economics Letters}\ }\textbf {\bibinfo
  {volume} {6}},\ \bibinfo {pages} {255 } (\bibinfo {year} {1980})}\BibitemShut
  {NoStop}%
\bibitem [{\citenamefont {D'Agostino}(1970)}]{DAgostino1970}%
  \BibitemOpen
  \bibfield  {author} {\bibinfo {author} {\bibfnamefont {R.~B.}\ \bibnamefont
  {D'Agostino}},\ }\href {http://www.jstor.org/stable/2334794} {\bibfield
  {journal} {\bibinfo  {journal} {Biometrika}\ }\textbf {\bibinfo {volume}
  {57}},\ \bibinfo {pages} {pp. 679} (\bibinfo {year} {1970})}\BibitemShut
  {NoStop}%
\bibitem [{\citenamefont {Aghanim}\ \emph {et~al.}(2015)\citenamefont {Aghanim}
  \emph {et~al.}}]{Aghanim:2015xee}%
  \BibitemOpen
  \bibfield  {author} {\bibinfo {author} {\bibfnamefont {N.}~\bibnamefont
  {Aghanim}} \emph {et~al.} (\bibinfo {collaboration} {Planck Collaboration}),\
  }\href@noop {} {\enquote {\bibinfo {title} {{Planck 2015 results. XI. CMB
  power spectra, likelihoods, and robustness of parameters}},}\ } (\bibinfo
  {year} {2015}),\ \bibinfo {note} {submitted to A\&A},\ \Eprint
  {http://arxiv.org/abs/1507.02704} {arXiv:1507.02704 [astro-ph.CO]}
  \BibitemShut {NoStop}%
\bibitem [{\citenamefont {Gorski}\ \emph {et~al.}(2005)\citenamefont {Gorski},
  \citenamefont {Hivon}, \citenamefont {Banday}, \citenamefont {Wandelt},
  \citenamefont {Hansen}, \citenamefont {Reinecke},\ and\ \citenamefont
  {Bartelman}}]{Gorski:2004by}%
  \BibitemOpen
  \bibfield  {author} {\bibinfo {author} {\bibfnamefont {K.~M.}\ \bibnamefont
  {Gorski}}, \bibinfo {author} {\bibfnamefont {E.}~\bibnamefont {Hivon}},
  \bibinfo {author} {\bibfnamefont {A.~J.}\ \bibnamefont {Banday}}, \bibinfo
  {author} {\bibfnamefont {B.~D.}\ \bibnamefont {Wandelt}}, \bibinfo {author}
  {\bibfnamefont {F.~K.}\ \bibnamefont {Hansen}}, \bibinfo {author}
  {\bibfnamefont {M.}~\bibnamefont {Reinecke}}, \ and\ \bibinfo {author}
  {\bibfnamefont {M.}~\bibnamefont {Bartelman}},\ }\href {\doibase
  10.1086/427976} {\bibfield  {journal} {\bibinfo  {journal} {Astrophys. J.}\
  }\textbf {\bibinfo {volume} {622}},\ \bibinfo {pages} {759} (\bibinfo {year}
  {2005})},\ \Eprint {http://arxiv.org/abs/astro-ph/0409513}
  {arXiv:astro-ph/0409513 [astro-ph]} \BibitemShut {NoStop}%
\end{thebibliography}%

\end{document}